\documentclass[prd,twocolumn,reprint,superscriptaddress,showpac]{revtex4-2}
\usepackage{feynmp,amsmath,amssymb,graphicx,subfigure,hyperref,bm,mathrsfs,slashed}
\usepackage{CJK}
\usepackage{color}

\begin{document}
\begin{CJK}{UTF8}{song}

\title{Signatures of Lifshitz transition in the optical conductivity of two-dimensional tilted Dirac materials}

\author{Chao-Yang Tan}
\thanks{These authors have contributed equally to this work.}
\affiliation{Department of Physics, Institute of Solid State Physics and
Center for Computational Sciences, Sichuan Normal University, Chengdu,
Sichuan 610066, China}
\affiliation{Department of Physics and Beijing Key Laboratory of Opto-electronic Functional
Materials and Micro-nano Devices, Renmin University of China, Beijing 100872, China}

\author{Jian-Tong Hou}
\thanks{These authors have contributed equally to this work.}
\affiliation{College of Physical Science and Technology Sichuan University, Chengdu,
Sichuan 610064, China}
\affiliation{Department of Physics, Institute of Solid State Physics and
Center for Computational Sciences, Sichuan Normal University, Chengdu,
Sichuan 610066, China}

\author{Chang-Xu Yan}
\affiliation{Department of Physics, Institute of Solid State Physics and
Center for Computational Sciences, Sichuan Normal University, Chengdu,
Sichuan 610066, China}

\author{Hong Guo}
\affiliation{Department of Physics, McGill University, Montreal, Quebec, Canada H3A 2T8}
\affiliation{Department of Physics, Institute of Solid State Physics and
Center for Computational Sciences, Sichuan Normal University, Chengdu,
Sichuan 610066, China}

\author{Hao-Ran Chang}
\thanks{Corresponding author:hrchang@mail.ustc.edu.cn}
\affiliation{Department of Physics, Institute of Solid State Physics and
Center for Computational Sciences, Sichuan Normal University, Chengdu,
Sichuan 610066, China}
\affiliation{College of Physical Science and Technology Sichuan University, Chengdu,
Sichuan 610064, China}
\affiliation{Department of Physics, McGill University, Montreal, Quebec H3A 2T8, Canada}

\date{\today}

\begin{abstract}
Lifshitz transition is a kind of topological phase transition in which the Fermi surface is
reconstructed. It can occur in the two-dimensional (2D) tilted Dirac materials when the energy
bands change between the type-I phase ($0<t<1$) and the type-II phase ($t>1$) through the
type-III phase ($t=1$), where different tilts are parametrized by the values of $t$. In order
to characterize the Lifshitz transition therein, we theoretically investigate the longitudinal
optical conductivities (LOCs) in type-I, type-II, and type-III Dirac materials within linear
response theory. In the undoped case, the LOCs are constants either independent of the tilt
parameter in both type-I and type-III phases or determined by the tilt parameter in the type-II
phase. In the doped case, the LOCs are anisotropic and possess two critical frequencies determined
by $\omega=\omega_1(t)$ and $\omega=\omega_2(t)$, which are also confirmed by the joint density
of state. The tilt parameter and chemical potential can be extracted from optical experiments
by measuring the positions of these two critical boundaries and their separation
$\Delta\omega(t)=\omega_2(t)-\omega_1(t)$. With increasing the tilting, the separation becomes larger
in the type-I phase whereas smaller in the type-II phase. The LOCs in the regime of large photon energy
are exactly the same as that in the undoped case. The type of 2D tilted Dirac bands can be determined
by the asymptotic background values, critical boundaries and their separation in the LOCs. These
can therefore be taken as signatures of Lifshitz transition therein. The results of this work
are expected to be qualitatively valid for a large number of 2D tilted Dirac materials, such as
8-\emph{Pmmn} borophene monolayer, $\alpha$-SnS$_2$, TaCoTe$_2$, TaIrTe$_4$, and $1T^\prime$
transition metal dichalcogenides, due to the underlying intrinsic similarities of 2D tilted Dirac bands.
\end{abstract}

\maketitle

\end{CJK}

%%%%%%%%%%%%%%%%%%%%%%%%%%%%%%%%
\section{Introduction\label{Sec:intro}}
%%%%%%%%%%%%%%%%%%%%%%%%%%%%%%%%

Graphene has triggered extremely active research in two-dimensional (2D) Dirac materials characterized
by linear and/or hyperbolic energy dispersions around Dirac points in momentum space \cite{Science2004,RMP2009},
such as $\alpha$-(BEDT-TTF)$_2$I$_3$ \cite{JPSJ2006}, silicene \cite{PRBSilicene2007,PRLSilicene2009,PRBSilicene2011,
PRLSilicene2011,EzawaPRLSilicene2012}, graphene under uniaxial strain \cite{ChoiPRB2010}, 8-$Pmmn$ borophene \cite{Zhou8Pmmn2014PRL,Science8Pmmn2015,PRB8PmmnRapid2016,PRB8Pmmn2016}, transition metal dichalcogenides \cite{PRLMoS2010,PRLMoS2012,Science2014}, partially hydrogenated graphene \cite{TingPRB2016}, $\alpha$-SnS$_2$
\cite{NPGMa2016}, TaCoTe$_2$ \cite{PRBYang2019}, and TaIrTe$_4$ \cite{PRBLu2020}. Among them, 2D tilted Dirac
materials host tilted dispersions along a certain direction of wave vector and have been attracting increasing
interests theoretically and experimentally \cite{JPSJ2006,ChoiPRB2010,Zhou8Pmmn2014PRL,Science8Pmmn2015,
PRB8PmmnRapid2016,PRB8Pmmn2016,Science2014,TingPRB2016,NPGMa2016,PRBYang2019,PRBLu2020}. They exhibit many
significant qualitative differences in physical behaviors compared to their untilted counterparts, including plasmons \cite{JPSJNishine2011,JPSJNishine2010,PRBIurov2017,PRBAgarwal2017,PRBJafari2018,PRBMojarro2022},
optical conductivities \cite{JPSJNishine2010,PRBVerma2017,PRBIurov2018,PRBHerrera2019,PRBGoerbig2019,
PRBIurov2020,PRBMoS22021,PRBJDOS2021,PRBZheng2021,PRBMojarro2022}, Weiss oscillation \cite{PRBIslam2017},
spectrum of superconducting excitations \cite{PRBLi2017}, Klein tunneling \cite{PRBSHZhangPRB2018,PRBNguyen2018,
CPBZhou2022}, Kondo effects \cite{PRBSun2018}, Ruderman-Kittel-Kasuya-Yosida (RKKY) interactions \cite{PRBPaul2019,JMMMZhang2019}, Hall effects \cite{RQWPRB2020R,PRRRostami2020}, thermoelectric effects \cite{PRBGhosh2020}, thermal currents \cite{APLSengupta2020},
valley filtering \cite{NanotechnologyZhai2021}, gravitomagnetic effects \cite{PRRJafari2020}, Andreev reflection \cite{PRBJafari2020}, Coulomb bound states \cite{PhysicaELv2021}, guided modes \cite{SRHartmann2022}, and
valley-dependent time evolution of coherent electron states \cite{PRBStegmann2022}.

Lifshitz transition is a kind of topological phase transition in which the Fermi surface is reconstructed
\cite{Lifshitz1960}, which is crucial for understanding the novel states of matter and physical properties
around the transition. It accounts for the huge magnetoresistance in black phosphorus \cite{PRLBlackPhosphorus2014}, 
superconductivity in iron-based superconductors \cite{NatPhyssuperconductivity2010, CPLsuperconductivity2012,PRLsuperconductivity2015,NatCommunsuperconductivity2017,SciAdvsuperconductivity2017},
and abnormal transport behavior in heavy fermion materials and topological quantum materials \cite{PRL1162016,
PRL1172016,NatComun2017,CPM2017}. The Lifshitz transition occurs in tilted Dirac materials when the energy bands
change between the type-I phase (under-tilted, $0<t<1$) and the type-II phase (over-tilted, $t>1$) through the
type-III phase (critical-tilted, $t=1$) where $t$ is the tilt parameter \cite{Volovik2017,Volovik2018}. In
three-dimensional (3D) tilted Dirac bands, many physical properties can characterize the Lifshitz transition,
such as spin susceptibilities \cite{PRBHonerkamp2016}, magnetic response \cite{PRLGoerbig2016,PRBGoerbig2017},
Hall conductivity \cite{JETPLZyuzin2016}, anomalous Nernst effect \cite{PRBZyuzin2017,EPJBTewari2018},
magnetoresponse \cite{PRLYang2016}, Andreev reflection \cite{PRBSun2017}, Klein tunneling \cite{PRLBeenakker2016},
Kerr rotation \cite{PRBSonowal2019}, magnetothermal transport \cite{PRBAgarwal2019}, plasmon \cite{PRLAgarwal2020},
RKKY interaction \cite{RQWPRB2019}, and optical response \cite{PRBCarbotte2016,PRBCarbotte2017,PRBOrnigotti2022}.

Similarly, the Lifshitz transition can be realized in 2D tilted Dirac materials. For example, 8-\emph{Pmmn} borophene 
undergoes the Lifshitz transition under the control of tunable vertical electrostatic field \cite{PRB8pmmn2019}; the 
different compounds in $1T^\prime$ transition metal dichalcogenides correspond to different phases of the Lifshitz 
transition \cite{Science2014}: type-I phase ($1T^\prime$-$\mathrm{MoS}_2$ and $1T^\prime$-$\mathrm{MoSe}_2$), type-III 
phase ($1T^\prime$-$\mathrm{WSe}_2$), and type-II phase ($1T^\prime$-$\mathrm{MoTe}_2$ and $1T^\prime$-$\mathrm{WTe}_2$). 
Consequently, a very important issue is how to most effectively characterize the Lifshitz transition in 2D tilted Dirac 
materials. In 2D tilted Dirac bands, some indicators were already proposed to characterize the Lifshitz transition, 
including the spectrum of superconducting excitations \cite{PRBLi2017}, Coulomb bound states \cite{PhysicaELv2021}, and 
nonlinear optical response \cite{PRBOrnigotti2022}.

\begin{figure*}[htbp]
\centering
\includegraphics[width=16cm]{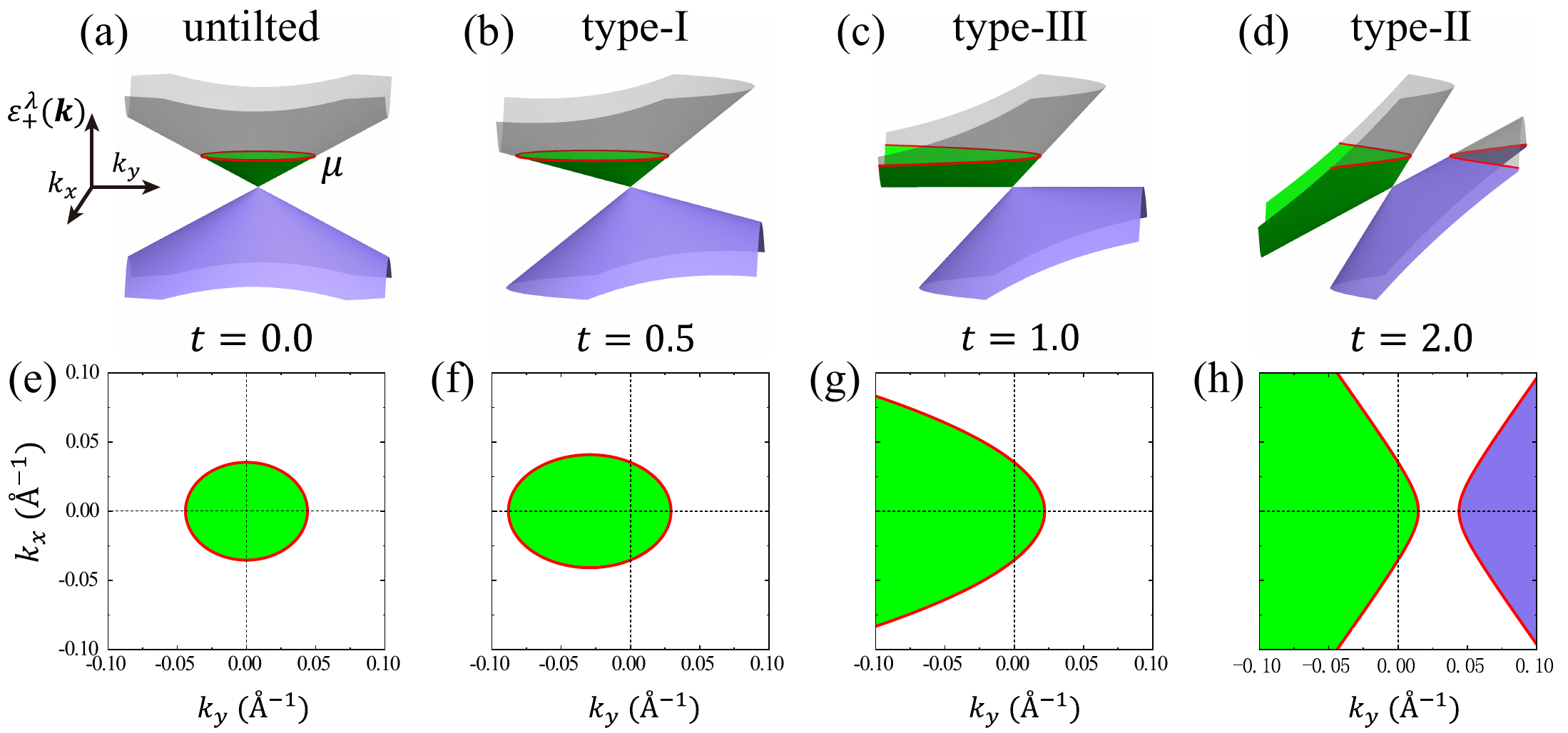}\\
\caption{Schematic diagrams for tilted Dirac bands (top panels) and the corresponding Fermi surfaces (bottom panels)
of different typical tilt parameters $t$ at the $\kappa=+$ valley. The green and purple shaded regions represent electron
and hole pockets, respectively. The Fermi contour represented by the red line, is an ellipse for $t=0$ and $t=0.5$,
a parabola for $t=1$, and a couple of hyperbola for $t=2$. We hereafter set the chemical potential $\mu=0.2$ eV and
the Fermi velocities $v_x=0.86v_F$, $v_y=0.69v_F$ with $v_F=10^6$ m/s, the same as that given in Ref. \cite{PRBVerma2017}.}
\label{fig1}
\end{figure*}

The optical conductivity provides a powerful method for extracting the information of energy band structure,
and has been extensively investigated theoretically and experimentally in the 2D \emph{untilted} Dirac bands
\cite{PRLCarbotte2006,PRBGusynin2007,PRLMikhailov2007,PRLMarel2008,PRLMak2008,PRBStauber2008,NatLi2008,PRBStille2012,
PRBCarbotte2012,PRBCarvalho2013,PRBAsgari2014,PRBDiPietro2012,PRBCarbotte2013,PRBXiao2013} and under-tilted Dirac bands \cite{JPSJNishine2010,PRBVerma2017,PRBHerrera2019,PRBGoerbig2019,PRBMoS22021,PRBJDOS2021,PRBMojarro2022}. Specifically,
the exotic behaviors of longitudinal optical conductivity (LOC) can be used to characterize the topological phase
transitions in both silicene \cite{PRBStille2012} and $1T^\prime$-$\mathrm{MoS}_2$ \cite{PRBMoS22021}. However, the
energy bands of the above-mentioned 2D Dirac materials are restricted to either the untilted or the under-tilted (type-I 
phase), leaving the impact of type-II and type-III energy bands on the LOCs unexplored. To characterize the Lifshitz
transition of tilted Dirac materials, we perform a comprehensive study of the LOCs in the type-I, type-II, and type-III
phases. In particular, we focus our theoretical study of LOCs on both the undoped and doped situations.

The rest of the paper is organized as follows. In Sec. \ref{Sec:Theoretical formalism}, we briefly describe the model Hamiltonian 
and theoretical formalism to calculate the LOC. The analytical expressions for the interband conductivity, joint density of states 
(JDOS), and the results for interband conductivity are presented in Sec. \ref{Sec:Interband conductivity} and Sec. \ref{Sec:Results}, 
respectively. In addition, the intraband conductivities are analytically calculated in Sec. \ref{Sec:Drude conductivity}. The summary 
and conclusions are given in Sec. \ref{Sec:CONCLUSIONS}. Finally, we present four appendices to show detailed calculations.

%%%%%%%%%%%%%%%%%%%%%%%%%%%%%%%%
\section{Theoretical formalism \label{Sec:Theoretical formalism}}
%%%%%%%%%%%%%%%%%%%%%%%%%%%%%%%%

We begin with the Hamiltonian in the vicinity of one of two valleys for 2D tilted Dirac materials \cite{JPSJ2006,ChoiPRB2010,Zhou8Pmmn2014PRL,Science8Pmmn2015,PRB8PmmnRapid2016,PRB8Pmmn2016,JPSJNishine2010}
\begin{align}
\mathcal{H}_\kappa(k_x,k_y)=\kappa \hbar v_t k_y\tau_0+\hbar (v_xk_x\tau_1+v_yk_y\tau_2),
\label{Eq1}
\end{align}
where $\kappa=\pm$ labels two valleys, $\mathcal{\textbf{\emph{k}}}=(k_x,k_y)$ stands for the wave vector,
and $\tau_0$ and $\tau_i$ denote the $2\times2$ unit matrix and Pauli matrices, respectively. For simplicity,
we hereafter set $\hbar=1$ and introduce the tilt parameter $t$ by defining
\begin{align}
t=\frac{v_t}{v_y}.
\end{align}
It is noted that this system remains invariant under the valley transformation $(\kappa,k_y)\leftrightarrow(-\kappa,-k_y)$, indicating $\mathcal{H}_{-\kappa}(k_x,-k_y)=\mathcal{H}_\kappa(k_x,k_y)$. The eigenvalue evaluated from the Hamiltonian
reads
\begin{align}
\varepsilon_\kappa^\lambda(k_x,k_y)=\kappa t v_y k_y+\lambda \mathcal{Z}(k_x,k_y),
\label{Eq2}
\end{align}
where $\mathcal{Z}(k_x,k_y)=\sqrt{v_x^2 k_x^2+v_y^2 k_y^2}$, and $\lambda=\pm1$ denotes the conduction band ($\lambda=+1$)
and valence band ($\lambda=-1$), respectively. The energy bands and the Fermi surfaces for the $n$-doped case at the
$\kappa=+$ valley, are schematically shown in Fig. \ref{fig1}.

For the untilted case ($t=0$), the only Fermi surface,
contributed completely by the electron pocket with a closed area [see Figs. \ref{fig1}(a) and \ref{fig1}(e)], is an
ellipse obeying the equation
\begin{align}
\frac{k_x^2}{\frac{\mu^2}{v_x^2}}+\frac{k_y^2}{\frac{\mu^2}{v_y^2}} =1,
\label{fermisurface1}
\end{align}
which reduces to a circle when $v_x=v_y$. For the type-I phase ($0<t<1$), the only Fermi surface, also contributed
completely by the closed electron pocket [see Figs. \ref{fig1}(b) and \ref{fig1}(f)], is an ellipse obeying the equation
\begin{align}
\frac{k_x^2}{\frac{\mu^2}{v_x^2(1-t^2)}}+\frac{\left(k_y+\frac{\kappa t\mu}{v_y(1-t^2)}\right)^2}{\frac{\mu^2}{v_y^2(1-t^2)^2}} =1,
\label{fermisurface2}
\end{align}
which remains an ellipse even when $v_x=v_y$. The tilt parameter $t$ moves the center of the ellipse along the $k_y$ axis
and changes the major axis and minor axis of the ellipse. For the type-III phase ($t=1$), the only Fermi surface, contributed entirely by the electron pocket with an open border [see Figs. \ref{fig1}(c) and \ref{fig1}(g)], is a parabola satisfying
\begin{align}
k_y=\frac{\mu}{2\kappa v_y}-\frac{v_x^2k_x^2}{2\kappa\mu v_y}.
\label{fermisurface3}
\end{align}
Interestingly, for the type-II phase ($t>1$), the Fermi surface is a couple of hyperbola [see Figs. \ref{fig1}(d) and \ref{fig1}(h)] whose equation reads
\begin{align}
\frac{\left(k_y-\frac{\kappa t \mu}{v_y(t^2-1)}\right)^2}{\frac{\mu^2}{v_y^2(t^2-1)^2}}
-\frac{k_x^2}{\frac{\mu^2}{v_x^2(t^2-1)}} =1,
\label{fermisurface4}
\end{align}
which is contributed not only by the electron pocket but also by the hole pocket.

These indicate that the Fermi surface is reconstructed when the energy band changes between the type-I phase and the
type-II phase, corresponding to a Lifshitz transition [see Figs. \ref{fig1}(e)$-$\ref{fig1}(h)]. On the other hand, the
edges of the electron pocket and the hole pocket determine the boundaries of the interband transition of the LOCs. As a consequence,
the Lifshitz transition can be characterized by the critical boundaries of interband conductivity. It is the purpose of
this work to characterize such Lifshitz transition in the 2D tilted Dirac materials via the LOC.

Within linear response theory, the LOC $\sigma_{jj}(\omega)$ at finite photon
frequency $\omega$ is given by
\begin{align}
\sigma_{jj}(\omega)&=g_s\sum_{\kappa=\pm1}\sigma_{jj}^\kappa(\omega),
\label{Eq3}
\end{align}
where $j=x,y$ stands for the spatial component, $g_s=2$ represents the spin degeneracy, and $\sigma_{jj}^\kappa(\omega)$
denotes the LOC at given valley $\kappa$, whose explicit expression is provided in the Appendix \ref{Sec:AppendixA}. Interestingly, $\sigma_{jj}(\omega)$ possesses the particle-hole symmetry (see Appendix \ref{Sec:AppendixA} for details)
such that we can safely replace $\mu$ by $|\mu|$ in all of $\sigma_{jj}(\omega)$, $\sigma_{jj}^{\kappa}(\omega)$,
and $f(x)$ because we only concern the final result of $\sigma_{jj}(\omega)$. It can be proven that $\sigma^{\kappa}_{jj}(\omega)=\sigma^{-\kappa}_{jj}(\omega)$ by considering $\mathcal{H}_{-\kappa}(k_x,-k_y)
=\mathcal{H}_\kappa(k_x,k_y)$, such that we are allowed to focus on the $\kappa=+$ or the $\kappa=-$ valley. Hereafter,
we restrict our analysis to the $n$-doped case ($\mu>0$) and the $\kappa=+$ valley for convenience.

\begin{widetext}
After some standard algebra, the real part of the LOCs can be divided into an interband part and an intraband part as
\begin{align}
\mathrm{Re}\sigma_{jj}^{\kappa}(\omega)
=\begin{cases}
\mathrm{Re}\sigma_{jj(\mathrm{IB})}^{\kappa}(\omega)+\Theta[\mu]\mathrm{Re}\sigma_{jj(\mathrm{D})}^{\kappa,+}(\omega)
+\Theta[-\mu]\mathrm{Re}\sigma_{jj(\mathrm{D})}^{\kappa,-}(\omega), & 0\leq t \leq1,\\\\
\mathrm{Re}\sigma_{jj(\mathrm{IB})}^{\kappa}(\omega)+\mathrm{Re}\sigma_{jj(\mathrm{D})}^{\kappa,+}(\omega)
+\mathrm{Re}\sigma_{jj(\mathrm{D})}^{\kappa,-}(\omega), & t>1,
\end{cases}
\label{Eq4}
\end{align}
where $\Theta(x)$ is the Heaviside step function satisfying $\Theta(x)=0$ for $x\le0$ and $\Theta(x)=1$ for $x>0$, and $\mu$ denotes the chemical potential measured with respect to the Dirac point. In addition, the interband and
intraband conductivities are given, respectively, as
\begin{align}
\mathrm{Re}\sigma_{jj(\mathrm{IB})}^\kappa(\omega)
&=\pi \int^{+\infty}_{-\infty}\frac{dk_x}{2\pi} \int^{+\infty}_{-\infty}\frac{dk_y}{2\pi}
\mathcal{F}^{\kappa;jj}_{-,+}(k_x,k_y) \frac{f\left[\varepsilon_\kappa^{-}(k_x,k_y)\right] -f\left[\varepsilon_\kappa^{+}(k_x,k_y)\right]} {\omega} \delta\left[\omega-2\mathcal{Z}(k_x,k_y)\right],\label{Eq5}\\
\mathrm{Re}\sigma_{jj(\mathrm{D})}^{\kappa,\lambda}(\omega)
&=\pi\int^{+\infty}_{-\infty}\frac{dk_x}{2\pi} \int^{+\infty}_{-\infty}\frac{dk_y}{2\pi} \mathcal{F}^{\kappa;jj}_{\lambda,\lambda}(k_x,k_y) \left[-\frac{d f\left[\varepsilon_\kappa^{\lambda}(k_x,k_y)\right]} {d\varepsilon_\kappa^{\lambda}(k_x,k_y)}\right]\delta(\omega),
\label{Eq6}
\end{align}
where $\delta(x)$ is the Dirac $\delta$-function, $f(x)=\left\{1+\exp[(x-\mu)/k_BT]\right\}^{-1}$ denotes the
Fermi distribution function in which $k_B$ is the Boltzmann constant and $T$ represents the temperature, and 
$\mathcal{F}_{\lambda,\lambda^\prime}^{\kappa;jj}(k_x,k_y)$ is explicitly given as
\begin{align}
\mathcal{F}_{\lambda,\lambda^\prime}^{\kappa;xx}(k_x,k_y)
=&\frac{e^2}{2}v_x^2 \left\{1+\lambda\lambda^\prime\frac{v_x^2k_x^2-v_y^2k_y^2} {\left[\mathcal{Z}(k_x,k_y)\right]^2}\right\},\\
\mathcal{F}_{\lambda,\lambda^\prime}^{\kappa;yy} (k_x,k_y)
=&\frac{e^2}{2}v_y^2\left\{2t^2\delta_{\lambda\lambda^\prime}+
1-\lambda\lambda^\prime\frac{v_x^2k_x^2-v_y^2k_y^2}{\left[\mathcal{Z}(k_x,k_y)\right]^2}
+4\lambda\delta_{\lambda\lambda^\prime}\frac{\kappa t v_y k_y}{\mathcal{Z}(k_x,k_y)}\right\},
\end{align}
with $\delta_{\lambda\lambda^\prime}$ the Kronecker symbol.

\end{widetext}

For the sake of simplicity, we denote the real part of total LOCs as
\begin{align}
\mathrm{Re}\sigma_{jj}(\omega)&=\mathrm{Re}\sigma_{jj}^{\mathrm{IB}}(\omega)+\mathrm{Re}\sigma_{jj}^{\mathrm{D}}(\omega),
\label{Eq7}
\end{align}
where $\mathrm{Re}\sigma_{jj}^{\mathrm{IB}}(\omega)$ and $\mathrm{Re}\sigma_{jj}^{\mathrm{D}}(\omega)$ are recast as
\begin{align}
\frac{\mathrm{Re}\sigma_{jj}^{\mathrm{IB}}(\omega)}{\sigma_0}&=\frac{v_x}{v_y}\Gamma_{xx}^{\mathrm{IB}}(\omega)\delta_{jx} +\frac{v_y}{v_x}\Gamma_{yy}^{\mathrm{IB}}(\omega)\delta_{jy},\label{Eq7IB}\\
\frac{\mathrm{Re}\sigma_{jj}^{\mathrm{D}}(\omega)}{\sigma_0}&
=\frac{v_x}{v_y}\Gamma_{xx}^{\mathrm{D}}(\mu,t,\Lambda)\delta_{jx}~\delta(\omega)
\nonumber\\&
+\frac{v_y}{v_x}\Gamma_{yy}^{\mathrm{D}}(\mu,t,\Lambda)\delta_{jy}~\delta(\omega).\label{Eq7D}
\end{align}
In these notations, $\sigma_0=e^2/4\hbar$ (we temporarily restore $\hbar$ for explicitness), and $\Gamma^{\mathrm{IB}}_{jj}(\omega)$ and $\Gamma^{\mathrm{D}}_{jj}(\mu,t,\Lambda)$ are introduced for convenience with $j=x,y$, whose explicit definitions are
independent of the ratio $v_x/v_y$ (see Appendix \ref{Sec:AppendixB} for details). The relation between the LOCs and
the ratio of Fermi velocities $v_x/v_y$ has also been reported in Refs. \cite{PRBHerrera2019} and \cite{PRBMoS22021}.
Obviously, the only difference between the \emph{isotropic} case ($v_x=v_y$) and the \emph{anisotropic} case ($v_x\neq v_y$)
is the magnitude of the LOCs. It is emphasized that $\sigma_0\frac{v_x}{v_y}\Gamma_{xx}^{\mathrm{D}}(\mu,t,\Lambda)$ and $\sigma_0\frac{v_y}{v_x}\Gamma_{yy}^{\mathrm{D}}(\mu,t,\Lambda)$ refer to the Drude weights for $\mathrm{Re}\sigma_{xx}^{\mathrm{D}}$ and $\mathrm{Re}\sigma_{yy}^{\mathrm{D}}$, respectively.

In the next, we will analytically calculate the interband and intraband LOCs by assuming zero temperature $T=0$ such
that the Fermi distribution function $f(x)$ can be replaced by the Heaviside step function $\Theta[\mu-x]$. To better
analyze the physics of interband LOCs, we also evaluate the joint density of states JDOS defined by
\begin{align}
\mathcal{J}(\omega)=g_s\sum\limits_{\kappa=\pm1}\mathcal{J}_{\kappa}(\omega),
\end{align}
where
\begin{align}
\mathcal{J}_{\kappa}(\omega)
&=\int\frac{d^2\boldsymbol{k}}{(2\pi)^2}
\delta\left[\omega-2\mathcal{Z}(k_x,k_y)\right]\notag\\
&\times\left\{\Theta\left[\mu-\varepsilon_\kappa^-(k_x,k_y)\right]
-\Theta\left[\mu-\varepsilon_\kappa^+(k_x,k_y)\right]\right\}.
\end{align}
The detailed analytical calculations of LOCs and JDOS are found in Appendices \ref{Sec:AppendixB} and \ref{Sec:AppendixC}.

%%%%%%%%%%%%%%%%%%%%%%%%%%%%%%%%
\section{Interband conductivity and JDOS \label{Sec:Interband conductivity}}
%%%%%%%%%%%%%%%%%%%%%%%%%%%%%%%%

In this section, the analytical results of the interband LOC and JDOS are listed for different tilts. Firstly, the
LOCs in the undoped case ($\mu=0$) are completely contributed by the interband transition, which are given as
\begin{align}
\Gamma_{xx}^{\mathrm{IB}}(\omega)
&=\begin{cases}
1, & 0\le t \leq1\\\\
G_{-}\left(\frac{1}{t}\right)-G_{-}\left(-\frac{1}{t}\right), & t>1
\end{cases}
\end{align}
and
\begin{align}
\Gamma_{yy}^{\mathrm{IB}}(\omega)
&=\begin{cases}
1, & 0\le t \leq1\\\\
G_{+}\left(\frac{1}{t}\right)-G_{+}\left(-\frac{1}{t}\right), & t>1,
\end{cases}
\end{align}
where two auxiliary functions
\begin{align}
G_\pm(x)=\frac{1}{2}+\frac{\arcsin x}{\pi}\pm\frac{x\sqrt{1-x^2}}{\pi}
\end{align}
are introduced for simplicity. It is interesting to note that in the undoped case the LOCs are constant in
frequency. Specifically, these constant conductivities depend on the tilt parameter only in the type-II Dirac
materials, but are independent of the tilt parameter in both the type-I and type-III Dirac materials.

Hereafter, we focus on the interband LOCs in the doped case for different tilts. In the untilted case ($ t =0$),
we recover the result of the ordinary Dirac cone and get
\begin{align}
\Gamma_{xx}^{\mathrm{IB}}(\omega)&=\Gamma_{yy}^{\mathrm{IB}}(\omega)=\Theta(\omega-2\mu ),\notag
\end{align}
which are consistent with Refs. \cite{PRLCarbotte2006,PRLMikhailov2007,PRLMarel2008,PRLMak2008,PRBStauber2008}.
The corresponding JDOS is given as
\begin{align}
\frac{\mathcal{J}(\omega)}{\mathcal{J}_0\omega}=
\begin{cases}
0, &0<\omega<2\mu \\\\
1, &\omega\geq2\mu ,
\end{cases}
\end{align}
with $\mathcal{J}_0=\frac{1}{2\pi v_x v_y}$.

When the Dirac cone is tilted ($t>0$), we introduce three compacted notations
\begin{align}
\xi_{\pm}&=\frac{2\mu \pm\omega}{\omega}\frac{\Theta(t)}{t},\notag\\
\omega_1(t)=&2\mu \frac{\Theta(t)}{1+t}, \notag\\
\omega_2(t)
=&2\mu \Theta(t)\left[\frac{\Theta(1-t)}{1-t}+\frac{\Theta(t-1)}{t-1}\right],\notag
\end{align}
in order to simplify our results. The reason that we still used the step function $\Theta(t)$ in these notations is just to emphasize 
the constraint $t>0$ without the need of referring to the context. In the subsequent three subsections, we list the analytical expressions 
of the interband conductivity for the type-I, type-II, and type-III Dirac materials in sequence. In the fourth subsection, we express $\Gamma_{xx}^{\mathrm{IB}}(\omega)$ and $\Gamma_{yy}^{\mathrm{IB}}(\omega)$ in terms of the corresponding JDOS.

%%%%%%%%%%%%%%%%%%%%%%%%%%%%%%%%
\subsection{For type-I Dirac materials}
%%%%%%%%%%%%%%%%%%%%%%%%%%%%%%%%

For the type-I phase ($0<t<1$), the interband conductivities can be expressed by
\begin{align}
\Gamma_{xx}^{\mathrm{IB}}(\omega)
&=\begin{cases}
0, & 0<\omega<\omega_1(t)\\\\
1-G_{-}(\xi_{-}), & \omega_1(t)\leq\omega<\omega_2(t)\\\\
1, & \omega\geq\omega_2(t),
\end{cases}
\label{Eq16}
\end{align}
and
\begin{align}
\Gamma_{yy}^{\mathrm{IB}}(\omega)
&=\begin{cases}
0, & 0<\omega<\omega_1(t)\\\\
1-G_{+}(\xi_{-}), & \omega_1(t)\leq\omega<\omega_2(t)\\\\
1, & \omega\geq\omega_2(t),
\end{cases}
\label{Eq17}
\end{align}
where $\xi_{\pm}=(2\mu \pm\omega)/ t \omega$. The corresponding JDOS $\mathcal{J}(\omega)$ is given by
\begin{align}
\frac{\mathcal{J}(\omega)}{\mathcal{J}_0\omega}=&
\begin{cases}
0, &0<\omega<\omega_1(t)\\\\
\frac{\arccos\xi_{-}}{\pi}, &\omega_1(t)\le\omega<\omega_2(t)\\\\
1, &\omega\ge\omega_2(t).
\end{cases}
\end{align}

It is noted that there are two tilt-dependent critical boundaries at $\omega=\omega_1(t)=2\mu /(1+t)$ and $\omega=\omega_2(t)=2\mu /(1-t)$ in the interband LOCs, which are also confirmed by the JDOS. In addition, $\frac{d\Gamma_{yy}^{\mathrm{IB}}(\omega)}{d\omega}$ is continuous at $\omega=\omega_1(t)$ and $\omega=\omega_2(t)$,
while $\frac{d\Gamma_{xx}^{\mathrm{IB}}(\omega)}{d\omega}$ is discontinuous thereabouts. These expressions agree
exactly with the analytical results in Ref. \cite{PRBMoS22021}. After substituting $v_x=0.86v_F$, $v_y=0.69v_F$
and $v_t=0.32v_F$ with $v_F=10^6$ m/s into Eq. (\ref{Eq7IB}), these expressions give rise to the numerical results
reported in Ref. \cite{PRBVerma2017}. Furthermore, these results are also valid in the untilted limit ($t\to0^{+}$)
and/or undoped case ($\mu=0$). In the regime of large photon energy where $\omega\gg\mathrm{Max}\{\omega_2(t),2\mu\}$
which leads to $\xi_{\pm}=\pm1/t$, these interband conductivities approach the asymptotic background values $\mathrm{Re}\sigma_{xx}^{\mathrm{asymp}}(\omega)=\frac{v_x}{v_y}\sigma_0$ and $\mathrm{Re}\sigma_{yy}^{\mathrm{asymp}}
(\omega)=\frac{v_y}{v_x}\sigma_0$, which satisfy $\mathrm{Re}\sigma_{xx}^{\mathrm{asymp}}(\omega)\times
\mathrm{Re}\sigma_{yy}^{\mathrm{asymp}}(\omega)=\sigma_0^2$. It is evident that the asymptotic background values and
their product are independent of the tilt parameter, which is the same as reported in Ref. \cite{PRBMoS22021}.

\begin{figure*}[htbp]
\centering
\includegraphics[width=16cm]{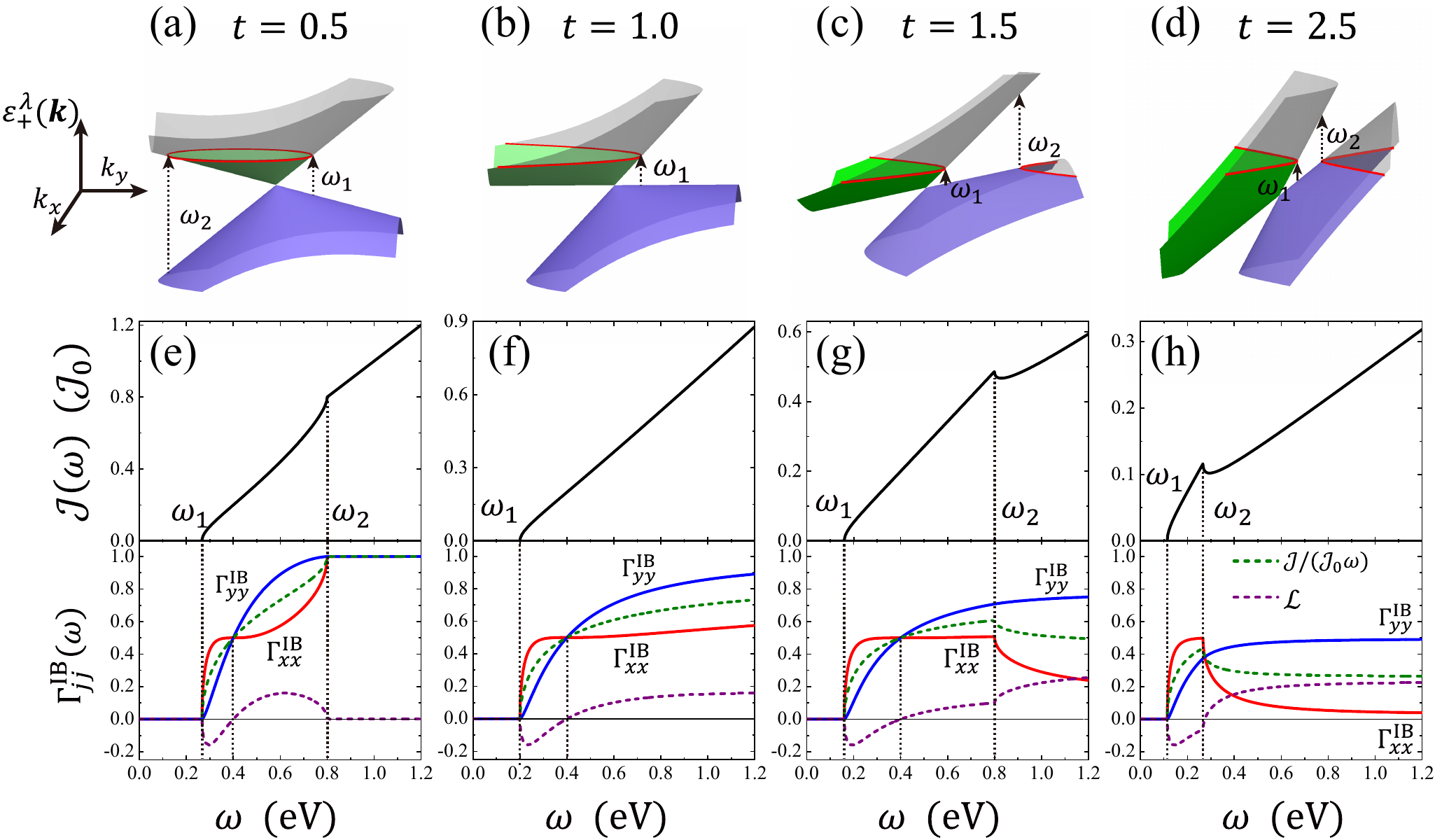}
\caption{Schematic interband transitions, the JDOS $\mathcal{J}(\omega)$, and the relation among $\Gamma_{jj}^{\mathrm{IB}}(\omega)$, $\mathcal{J}(\omega)$, and $\mathcal{L}(\omega)$. In panels (a),
(c), and (d), two Van Hove singularities appear at $\omega=\omega_{1}(t)$ and $\omega=\omega_{2}(t)$
in the type-I and type-II phases, but in panel (b), only one Van Hove singularity exists at
$\omega=\omega_{1}(t)$ in the type-III phase. In the top panel of (e)$-$(h), the JDOS $\mathcal{J}(\omega)$
and the corresponding Van Hove singularities are shown, while in the bottom panel of (e)$-$(h), $\Gamma_{jj}^{\mathrm{IB}}(\omega)$ and the relation with $\mathcal{J}(\omega)$ and
$\mathcal{L}(\omega)$ are presented, respectively.}
\label{fig2}
\end{figure*}

%%%%%%%%%%%%%%%%%%%%%%%%%%%%%%%%
\subsection{For type-II Dirac materials}
%%%%%%%%%%%%%%%%%%%%%%%%%%%%%%%%

For the type-II phase ($t>1$), the interband conductivities take the form
\begin{align}
\Gamma_{xx}^{\mathrm{IB}}(\omega)
=\begin{cases}
0, & 0<\omega<\omega_1(t)\\\\
1-G_{-}(\xi_{-}), & \omega_1(t)\leq\omega<\omega_2(t)\\\\
\sum\limits_{\chi=\pm1}\chi G_{-}(\xi_{\chi}), & \omega\geq\omega_2(t)
\end{cases}
\label{Eq18}
\end{align}
and
\begin{align}
\Gamma_{yy}^{\mathrm{IB}}(\omega)
=\begin{cases}
0, & 0<\omega<\omega_1(t)\\\\
1-G_{+}(\xi_{-}), & \omega_1(t)\leq\omega<\omega_2(t)\\\\
\sum\limits_{\chi=\pm1}\chi G_{+}(\xi_{\chi}), & \omega\geq\omega_2(t),
\end{cases}
\label{Eq19}
\end{align}
where $\xi_{\pm}=(2\mu \pm\omega)/t\omega$. The corresponding JDOS $\mathcal{J}(\omega)$ reads
\begin{align}
\frac{\mathcal{J}(\omega)}{\mathcal{J}_0\omega}=&
\begin{cases}
0, &0<\omega<\omega_1(t)\\\\
\frac{\arccos\xi_{-}}{\pi}, &\omega_1(t)\le\omega<\omega_2(t)\\\\
\frac{\arcsin\xi_{+}-\arcsin\xi_{-}}{\pi}, &\omega\ge\omega_2(t).
\end{cases}
\end{align}

There are also two tilt-dependent critical boundaries at $\omega=\omega_1(t)=2\mu /(t+1)$ and
$\omega=\omega_2(t)=2\mu /(t-1)$ in the interband LOCs, which are also confirmed by the JDOS.
Note that in this case, $\frac{d\Gamma_{yy}^{\mathrm{IB}}(\omega)}{d\omega}$ is continuous, 
while $\frac{d\Gamma_{xx}^{\mathrm{IB}}(\omega)}{d\omega}$ is discontinuous at these two tilt-dependent
critical boundaries. In the regime of large photon energy where $\omega\gg\mathrm{Max}\{\omega_2(t),2\mu \}$
which leads to $\xi_{\pm}=\pm1/t$, the asymptotic background values can be obtained as
\begin{align}
\mathrm{Re}\sigma_{xx}^{\rm asymp}(t)
=\frac{v_x}{v_y}\sigma_0\left[G_{-}\left(\frac{1}{t}\right)-G_{-}\left(-\frac{1}{t}\right)\right],\label{Eq20}\\
\mathrm{Re}\sigma_{yy}^{\rm asymp}(t)
=\frac{v_y}{v_x}\sigma_0\left[G_{+}\left(\frac{1}{t}\right)-G_{+}\left(-\frac{1}{t}\right)\right],\label{Eq21}
\end{align}
which is a straightforward consequence of the LOCs in the undoped case. In addition, they satisfy
\begin{align}
&\mathrm{Re}\sigma_{xx}^{\rm asymp}(t)\times\mathrm{Re}\sigma_{yy}^{\rm asymp}(t)
\notag\\
&=\frac{4}{\pi^2}\left[\frac{1}{t^4}-\frac{1}{t^2}+\arcsin^2\left(\frac{1}{t}\right)\right]\sigma_0^2,
\end{align}
which, different from that in the type-I phase, is tilt-dependent.

%%%%%%%%%%%%%%%%%%%%%%%%%%%%%%%%
\subsection{For type-III Dirac materials}
%%%%%%%%%%%%%%%%%%%%%%%%%%%%%%%%

For the type-III phase ($t=1$), the interband conductivities are given as
\begin{align}
\Gamma_{xx}^{\mathrm{IB}}(\omega)
=&\begin{cases}
0, & 0<\omega<\mu \\\\
1-G_{-}(\xi_{-}), & \omega\geq\mu
\end{cases}
\end{align}
and
\begin{align}
\Gamma_{yy}^{\mathrm{IB}}(\omega)
=&\begin{cases}
0, & 0<\omega<\mu \\\\
1-G_{+}(\xi_{-}), & \omega\geq\mu ,
\end{cases}
\end{align}
where $\xi_{\pm}=(2\mu \pm\omega)/\omega$. The corresponding JDOS $\mathcal{J}(\omega)$ is given by
\begin{align}
\frac{\mathcal{J}(\omega)}{\mathcal{J}_0\omega}
=&\begin{cases}
0, &0<\omega<\mu \\\\
\frac{\arccos\xi_{-}}{\pi}, &\omega\geq\mu .
\end{cases}
\end{align}

It is remarked that there is only one finite critical boundary at $\omega=\mu =\omega_1(1)$ in the interband LOCs,
which is also confirmed by the corresponding JDOS. Moreover, around this critical boundary, $\frac{d\Gamma_{yy}
^{\mathrm{IB}}(\omega)}{d\omega}$ is continuous, while $\frac{d\Gamma_{xx}^{\mathrm{IB}}(\omega)}{d\omega}$ is
not. In the regime of large photon energy where $\omega\gg2\mu $, these interband conductivities approach to
the asymptotic background values $\mathrm{Re}\sigma_{xx}^{\mathrm{asymp}}(\omega)=\frac{v_x}{v_y}\sigma_0$ and $\mathrm{Re}\sigma_{yy}^{\mathrm{asymp}}(\omega)=\frac{v_y}{v_x}\sigma_0$. The results for the interband LOCs,
the JDOS, the asymptotic background values, and their product $\mathrm{Re}\sigma_{xx}^{\mathrm{asymp}}(\omega)\times
\mathrm{Re}\sigma_{yy}^{\mathrm{asymp}}(\omega)$ can also be obtained from that of type-I phase in the limit $t\to1^{-}$
or from that of type-II phase in the limit $ t \to1^{+}$. As a consequence, the interband conductivities are continuous
when the tilt parameter changes from $t<1$ to $t>1$ through $t=1$.

\begin{figure*}[htbp]
\centering
\includegraphics[width=16cm]{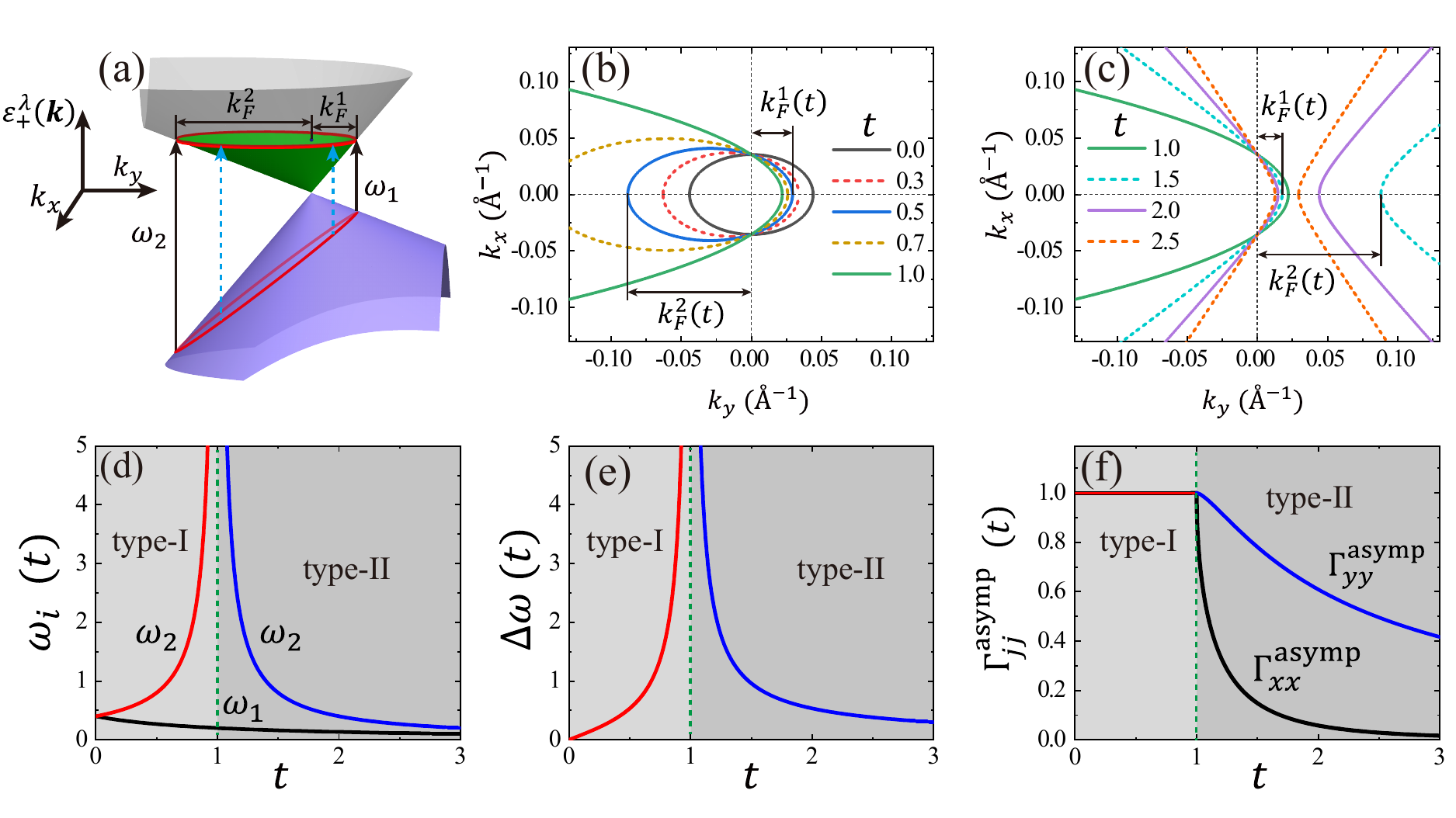}
\caption{Schematic diagram of interband transition, the dependence of two critical boundaries $\omega=\omega_i(t)$,
their separation $\Delta\omega(t)$, and the asymptotic background values $\Gamma_{jj}^{\mathrm{asymp}}(t)$ on the tilt
parameter $t$. In panel (a), the interband transitions and two critical boundaries are schematically shown, where
two critical boundaries $\omega=\omega_1(t)=2v_yk_F^1(t)$ and $\omega=\omega_2(t)=2v_yk_F^2(t)$ and the corresponding
Fermi wave vectors $k_F^1(t)=\frac{\mu}{(1+t)v_y}$ and $k_F^2(t)=\frac{\mu}{|1-t|v_y}$ are labeled. The Fermi surfaces
projected onto the $k_x$-$k_y$ plane are shown in panels (b) and (c). Two Fermi wave vectors $k_F^1(t)$ and $k_F^2(t)$
can reflect the corresponding changes in $\omega=\omega_1(t)$ and $\omega=\omega_2(t)$ with respect to $t$. As $t$ increases, $k_F^1(t)$ always decreases. In comparison, $k_F^2(t)$ becomes larger when $0<t<1$ but becomes smaller for $t>1$. In panel (d), the first critical boundary at $\omega=\omega_1(t)$ is represented by the black line in both the type-I and type-II phases, while
the second critical boundary at $\omega=\omega_2(t)$ is denoted either by the red line in the type-I phase or by the blue
line in the type-II phase. These are in exact agreement with the trend of $\omega_1(t)$ and $\omega_2(t)$ in panel (a)-(c).
The separation between two critical boundaries $\Delta\omega(t)$ is shown in (e): type-I phase (red line) and type-II
phase (blue line). The dependence of asymptotic background values $\Gamma_{jj}^{\mathrm{asymp}}(t)$ on the tilt parameter
$t$ are shown in (f): $\Gamma_{xx}^{\mathrm{asymp}}(t)=\Gamma_{yy}^{\mathrm{asymp}}(t)=1$ in the type-I phase but $\Gamma_{xx}^{\mathrm{asymp}}(t)<\Gamma_{yy}^{\mathrm{asymp}}(t)<1$ in the type-II phase.}
\label{fig3}
\end{figure*}

%%%%%%%%%%%%%%%%%%%%%%%%%%%%%%%%
\subsection{Relation between $\Gamma_{jj}^{\mathrm{IB}}(\omega)$ and $\mathcal{J}(\omega)$}
%%%%%%%%%%%%%%%%%%%%%%%%%%%%%%%%

In this subsection, we express $\Gamma_{xx}^{\mathrm{IB}}(\omega)$ and $\Gamma_{yy}^{\mathrm{IB}}(\omega)$ alternatively
in terms of the corresponding JDOS as
\begin{align}
\Gamma^{\mathrm{IB}}_{xx}(\omega) =\frac{\mathcal{J}(\omega)}{\mathcal{J}_0\omega}-\mathcal{L}(\omega),\label{Eq35}\\
%%%%%%%%%%%%%%%%%%%%%%%%%%%%%%%%%%%%%%%%%%%%%%
\Gamma^{\mathrm{IB}}_{yy}(\omega)
=\frac{\mathcal{J}(\omega)}{\mathcal{J}_0\omega}+\mathcal{L}(\omega).
\label{Eq36}
\end{align}

From these two relations, it can be found that $\Gamma_{xx}^{\mathrm{IB}}(\omega)$ and
$\Gamma_{yy}^{\mathrm{IB}}(\omega)$ depend on JDOS in terms of $\frac{\mathcal{J}(\omega)}{\mathcal{J}_0\omega}$
rather than $\mathcal{J}(\omega)$, and on the auxiliary function $\mathcal{L}(\omega)$ in the opposite way. The auxiliary
function $\mathcal{L}(\omega)$ is explicitly written as
\begin{align}
\mathcal{L}(\omega)=g_s\sum\limits_{\kappa=\pm1}\mathcal{L}_\kappa(\omega),
\end{align}
where
\begin{align}
\mathcal{L}_\kappa(\omega)&=2\pi v_xv_y
\int\frac{d^2\boldsymbol{k}}{(2\pi)^2}
\frac{v_x^2k_x^2-v_y^2k_y^2}{\left[\mathcal{Z}(k_x,k_y)\right]^2}\delta\left[\omega-2\mathcal{Z}(k_x,k_y)\right]
\nonumber\\&
\times\frac{\Theta\left[\mu-\varepsilon^-_{\kappa}(k_x,k_y)\right]
-\Theta\left[\mu-\varepsilon_\kappa^{+}(k_x,k_y)\right]}{\omega}.
\end{align}

It is emphasized that in the untilted phase ($t=0$), we have $\mathcal{L}_\kappa(\omega)=0$ after taking the symmetry of integration into account, and hence the auxiliary function $\mathcal{L}(\omega)=0$. As a direct result, the dimensionless function $\Gamma^{\mathrm{IB}}_{jj}(\omega)$ can be written as
$\Gamma^{\mathrm{IB}}_{xx}(\omega)=\Gamma^{\mathrm{IB}}_{yy}(\omega)
=\frac{\mathcal{J}(\omega)}{\mathcal{J}_0\omega}$.
However, in the tilted phase ($t>0$), the auxiliary function $\mathcal{L}(\omega)$ does not always vanish along
other directions. As a direct consequence, $\Gamma^{\mathrm{IB}}_{xx}(\omega)\neq\Gamma^{\mathrm{IB}}_{yy}(\omega)$,
which implies that $\mathcal{L}(\omega)$ can characterize the anisotropy between $\Gamma^{\mathrm{IB}}_{xx}(\omega)$
and $\Gamma^{\mathrm{IB}}_{yy}(\omega)$ originating from the tilting of Dirac bands.

After some straightforward algebra (see Appendix \ref{Sec:AppendixD} for details), we arrive at the explicit expressions of
$\mathcal{L}(\omega)$ for the type-I phase ($0<t<1$) as
\begin{align}
\mathcal{L}(\omega)
&=
\begin{cases}
0, & 0<\omega<\omega_1(t) \\\\
\frac{-\xi_{-}\sqrt{1-\xi_{-}^2}}{\pi}, & \omega_1(t)\leq\omega<\omega_2(t) \\\\
0, & \omega\geq\omega_2(t),
\end{cases}
\end{align}
for the type-II phase ($t>1$) as
\begin{align}
\mathcal{L}(\omega)
&=
\begin{cases}
0, & 0<\omega<\omega_1(t) \\\\
\frac{-\xi_{-}\sqrt{1-\xi_{-}^2}}{\pi}, & \omega_1(t)\leq\omega<\omega_2(t) \\\\
\frac{\xi_{+}\sqrt{1-\xi_{+}^2}-\xi_{-}\sqrt{1-\xi_{-}^2}}{\pi}, & \omega\geq\omega_2(t),
\end{cases}
\end{align}
and for the type-III phase ($t=1$) as
\begin{align}
\mathcal{L}(\omega)
&=
\begin{cases}
0, & 0<\omega<\mu \\\\
\frac{-\xi_{-}\sqrt{1-\xi_{-}^2}}{\pi}, & \omega\geq\mu.
\end{cases}
\end{align}

%%%%%%%%%%%%%%%%%%%%%%%%%%%%%%%
\section{RESULTS FOR THE INTERBAND CONDUCTIVITY\label{Sec:Results}}
%%%%%%%%%%%%%%%%%%%%%%%%%%%%%%%

Utilizing the analytical expressions listed in the previous section, we plot the interband transitions, JDOS,
and LOCs in Fig. \ref{fig2}. We at first present four general findings. First, the interband conductivities are
anisotropic, due to the band tilting and Pauli blocking. Second, when $t\neq0$ and $t\neq1$, the JDOS possesses two Van Hove singularities
at $\omega=\omega_1(t)$ and $\omega=\omega_2(t)$, leading to two critical boundaries in the LOC. Third, in the
region $0<\omega<\omega_1(t)$, the real part of LOCs always vanish, namely, $\Gamma_{jj}^{\mathrm{IB}}(\omega)\equiv0$. Fourth, the
results are valid for both the $n$-doped and $p$-doped cases.

For the type-I phase ($0<t<1$), it can be seen from Fig. \ref{fig2}(e) that in the region $\omega_1(t)\leq\omega<\omega_2(t)$, $\Gamma_{jj}^{\mathrm{IB}}(\omega)$ increases monotonically from zero at $\omega=\omega_1(t)$ to one at $\omega=\omega_2(t)$, and that $\Gamma_{jj}^{\mathrm{IB}}(\omega)\equiv1$ when $\omega\ge\omega_2(t)$. By contrast, for the type-II
phase ($t>1$), as shown in Figs. \ref{fig2}(g) and \ref{fig2}(h), $\Gamma_{jj}^{\mathrm{IB}}(\omega)$ increases monotonically
from zero at $\omega=\omega_1(t)$ to the corresponding critical value at $\omega=\omega_2(t)$. When $\omega\ge\omega_2(t)$, however, $\Gamma_{xx}^{\mathrm{IB}}(\omega)$ drops dramatically in magnitude whereas $\Gamma_{yy}^{\mathrm{IB}}(\omega)$
becomes larger smoothly with the increasing of photon energy. This difference results from the competition between $\frac{\mathcal{J}(\omega)}{\mathcal{J}_0\omega}$ and $\mathcal{L}(\omega)$. Explicitly, when $\omega\ge\omega_2(t)$, $\frac{\mathcal{J}(\omega)}{\mathcal{J}_0\omega}$ is identically equal to one in the type-I phase whereas $\frac{\mathcal{J}(\omega)}{\mathcal{J}_0\omega}$ decreases monotonously in the type-II phase, the latter of
which originates from the dip in JDOS. On the other hand, in this region $\mathcal{L}(\omega)$ vanishes for
the type-I phase whereas $\mathcal{L}(\omega)$ increases monotonously with $\omega$ for the type-II
phase. In the region $\omega>\omega_2(t)$, the relative changes of $\frac{\mathcal{J}(\omega)}{\mathcal{J}_0\omega}$
and $\mathcal{L}(\omega)$ determine the dramatically different behaviors of $\Gamma_{xx}^{\mathrm{IB}}(\omega)$
and $\Gamma_{yy}^{\mathrm{IB}}(\omega)$ in the type-I and type-II phases, leading to a consequence that
$\Gamma_{xx}^{\mathrm{IB}}(\omega)=\Gamma_{yy}^{\mathrm{IB}}(\omega)$ in the type-I phase while
$\Gamma_{xx}^{\mathrm{IB}}(\omega)\neq\Gamma_{yy}^{\mathrm{IB}}(\omega)$ in the type-II phase, as shown in the
bottom panels of Fig. \ref{fig2}. Specifically, for the type-III phase ($t=1$), the JDOS exhibits one Van Hove
singularity at $\omega=\omega_1(t)=\mu$ and one Van Hove singularity at $\omega=\omega_2(t)=\infty$. As a result,
there is only one critical boundary at finite frequency in the LOCs, as shown in Fig. \ref{fig2}(f). For the untilted
case ($t=0$), the JDOS exhibits only one Van Hove singularity at $\omega=2\mu$, so the LOC behaves
as a step function.

Interestingly, when $0<t\le2$,
\begin{align}
\Gamma_{xx}^{\mathrm{IB}}(2\mu)=\Gamma_{yy}^{\mathrm{IB}}(2\mu)=\frac{1}{2},
\end{align}
as a consequence of $\mathcal{L}(2\mu)=0$, and when $t\ge2$,
\begin{align}
&\Gamma_{xx}^{\mathrm{IB}}\left(2\mu\sqrt{2/(t^2-2)}\right)
=\Gamma_{yy}^{\mathrm{IB}}\left(2\mu\sqrt{2/(t^2-2)}\right)
\notag\\&
=\frac{\arcsin\left[\zeta_{+}(t)\right]-\arcsin\left[\zeta_{-}(t)\right]}{\pi}
\end{align}
with $\zeta_{\pm}(t)=\frac{1}{t}(\sqrt{\frac{t^2-2}{2}}\pm1)$.

The tilt parameter $t$ and the chemical potential $\mu $ can be extracted from optical experiments by measuring two
critical boundaries at $\omega=\omega_1(t)$ and $\omega=\omega_2(t)$. As shown in Fig. \ref{fig3}, the first critical
boundary $\omega=\omega_1(t)$ decreases monotonically with the increasing of $t$ no matter in the type-I phase ($0<t<1$)
or type-II phase ($t>1$), however, the second critical boundary $\omega=\omega_2(t)$ increases in the type-I phase ($0<t<1$)
and decreases in the type-II phase ($t>1$) as $t$ increases. Explicitly, the tilt parameter $t$ satisfies the relation
\begin{align}
t&=\frac{\omega_2(t)-\omega_1(t)}{\omega_2(t)+\omega_1(t)}.
\end{align}
Combined with the separation between two critical boundaries
\begin{align}
\Delta\omega(t)
=\omega_2(t)-\omega_1(t)=\begin{cases}
\frac{4\mu t}{1-t^2}, & 0<t<1\\\\
\frac{4\mu }{t^2-1}, & t>1,
\end{cases}
\end{align}
one can further determine the chemical potential $\mu$. Interestingly, it is shown in Fig. \ref{fig3}(e) that with
increasing the tilt parameter $t$ this separation becomes larger in the type-I phase ($0<t<1$) whereas smaller in the
type-II phase ($t>1$). In experiments, the positions of these two critical boundaries and their separation can also
be used to determine whether the Lifshitz transition occurs and which phase the Dirac materials belong to.

In addition, the asymptotic background values $\Gamma_{jj}^{\mathrm{asymp}}(t)$ behave dramatically different in the type-I
phase and type-II phase, as shown in Fig. \ref{fig3}(f). For the type-I phase, $\Gamma_{xx}^{\mathrm{asymp}}(t)$ and 
$\Gamma_{yy}^{\mathrm{asymp}}(t)$ are always identically equal to $1$. By contrast, for the type-II phase, $\Gamma_{xx}
^{\mathrm{asymp}}(t)$ and $\Gamma_{yy}^{\mathrm{asymp}}(t)$ decrease with $t$ and always satisfy the relation 
$\Gamma_{xx}^{\mathrm{asymp}}(t)<\Gamma_{yy}^{\mathrm{asymp}}(t)<1$. In addition, for the type-III phase, the asymptotic 
background values satisfy $\Gamma_{xx}^{\mathrm{asymp}}(t=1)=\Gamma_{yy}^{\mathrm{asymp}}(t=1)=1$. Hence, the asymptotic 
background values $\Gamma_{jj}^{\mathrm{asymp}}(t)$ can also be taken as fingerprints of type-I phase and type-II phase.

\begin{figure*}[htbp]
\centering
\includegraphics[width=12cm]{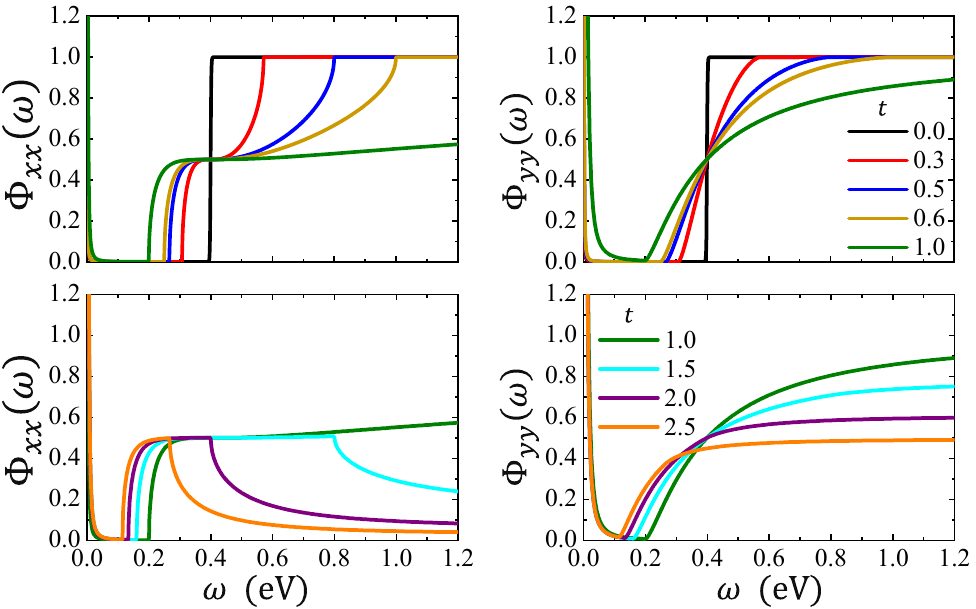}
\caption{The dependence of $\Phi_{xx}(\omega)$ and $\Phi_{yy}(\omega)$ on the tilt parameter. $\Phi_{xx}(\omega)$
and $\Phi_{yy}(\omega)$ are plotted by utilizing the relations given in Eqs. (\ref{Eq57})$-$(\ref{Eq59}). In the numerical
calculation, the chemical potential is set to $\mu=0.2$ eV.}
\label{fig4}
\end{figure*}

%%%%%%%%%%%%%%%%%%%%%%%%%%%%%%%
\section{Drude conductivity  \label{Sec:Drude conductivity}}
%%%%%%%%%%%%%%%%%%%%%%%%%%%%%%%

In this section, we turn to the Drude conductivities contributed by the intraband transition around the Fermi surface.
At zero temperature, the derivative of the Fermi distribution function in Eq. (\ref{Eq6}) can be replaced by $\delta[\mu -\varepsilon_\kappa^\lambda(k_x,k_y)]$. In the untilted case ($t=0$), we recover the result of the ordinary
Dirac cone as
\begin{align}
\Gamma_{xx}^{\mathrm{D}}(\mu,t,\Lambda)
&=\Gamma_{yy}^{\mathrm{D}}(\mu,t,\Lambda)
=4\mu,\notag
\end{align}
which is exactly the same as that in Ref. \cite{PRLCarbotte2006}. For the type-I phase ($0<t<1$), we arrive at
\begin{align}
\Gamma_{xx}^{\mathrm{D}}(\mu,t,\Lambda)
=&8\mu \frac{1-\sqrt{1-t^2}}{t^2\sqrt{1-t^2}},\\
\Gamma_{yy}^{\mathrm{D}}(\mu,t,\Lambda)
=&8\mu \frac{1-\sqrt{1-t^2}}{t^2}.
\end{align}

Four remarks for the type-I phase are in order here. First, these results are also valid in the untilted limit ($t\to0^{+}$).
Second, in the critical-tilted limit ($t\to1^{-}$), $\Gamma_{yy}^{\mathrm{D}}(\mu,t,\Lambda)=8\mu$ but $\Gamma_{xx}^{\mathrm{D}}(\mu,t,\Lambda)$ is divergent. Third, the ratio $\Gamma_{xx}^{\mathrm{D}}(\mu,t,\Lambda)/\Gamma_{yy}^{\mathrm{D}}(\mu,t,\Lambda)$ is always $1/\sqrt{1-t^2}$.
Fourth, $\Gamma_{jj}^{\mathrm{D}}(\mu,t,\Lambda)$ are always convergent when $0<t<1$, and yield the numerical
results of Drude weight in Ref. \cite{PRBVerma2017}, namely, $N_1=4.686$ and $N_2=2.673$ after applying the
parameters of 8-$Pmmn$ borophene ($v_x=0.86v_F$, $v_y=0.69v_F$, and $v_t=0.32v_F$ with $v_F=10^6$ m/s). The detailed evaluation can be found in Appendix \ref{Sec:AppendixB}.

For the type-II and type-III phases ($t\geq1$), we introduce a momentum cutoff $\Lambda$ to account for the limitation of integration interval, which is a measure of the density of states due to the electron and hole pockets \cite{PRBCarbotte2016}. For the type-II phase ($t>1$), the Drude contributions are given by
\begin{align}
\Gamma_{xx}^{\mathrm{D}}(\mu,t,\Lambda)
=&\frac{8\mu}{\pi} \Big[A(\mu ,t,\Lambda)\frac{2\Lambda}{\mu}
+\frac{B(\mu,t,\Lambda)}{\sqrt{t^2-1}}\nonumber\\
&\hspace{0.8cm}-C(\mu ,t,\Lambda)\Big],\\
\Gamma_{yy}^{\mathrm{D}}(\mu,t,\Lambda)
=&\frac{8\mu}{\pi}\Big[(t^2-1)A(\mu,t,\Lambda)\frac{2\Lambda}{\mu}\nonumber\\
&+\sqrt{t^2-1}B(\mu,t,\Lambda)+C(\mu ,t,\Lambda)\Big],
\end{align}
where
\begin{align}
A(\mu ,t,\Lambda)=&\sum_{\chi=\pm}\frac{1}{2t}\sqrt{1-\left(\frac{\mu -\chi\Lambda}{t\Lambda}\right)^2},\notag\\
B(\mu ,t,\Lambda)=&\frac{1}{t^2}\ln\frac{t^2+\frac{\mu -\Lambda}{\Lambda}
+\sqrt{t^2-1}\sqrt{t^2-(\frac{\mu -\Lambda}{\Lambda})^2}} {t^2-\frac{\mu +\Lambda}{\Lambda}
+\sqrt{t^2-1}\sqrt{t^2-(\frac{\mu +\Lambda}{\Lambda})^2}},\notag\\
C(\mu ,t,\Lambda)=&\sum_{\chi=\pm}\frac{1}{t^2}\arccos\frac{\mu -\chi\Lambda}{t\Lambda}.
\end{align}
Keeping up to $O(1)$ of $\Lambda$, we have the approximate expressions
\begin{align}
\Gamma_{xx}^{\mathrm{D}}(\mu,t,\Lambda)
&=8\mu\left[\frac{\sqrt{t^2-1}}{\pi t^2}\frac{2\Lambda}{\mu}-\frac{1}{t^2}\right],\\
\Gamma_{yy}^{\mathrm{D}}(\mu,t,\Lambda)
&=8\mu\left[\frac{\sqrt{(t^2-1)^3}}{\pi t^2}\frac{2\Lambda}{\mu}+\frac{1}{t^2}\right],
\end{align}
which indicate that both $\Gamma_{xx}^{\mathrm{D}}(\mu,t,\Lambda)$ and $\Gamma_{yy}^{\mathrm{D}}(\mu,t,\Lambda)$ are divergent when the cutoff $\Lambda$ is taken to be infinity.

For the type-III phase ($t=1$), the intraband contributions read
\begin{align}
\Gamma_{xx}^{\mathrm{D}}(\mu,t,\Lambda)
&=\frac{16\mu}{\pi}\left[\sqrt{\frac{2\Lambda-\mu }{\mu }}-\arccos\sqrt{\frac{\mu}{2\Lambda}}\right],\\
\Gamma_{yy}^{\mathrm{D}}(\mu,t,\Lambda)
&=\frac{8\mu}{\pi}\arccos\left(\frac{\mu-\Lambda}{\Lambda}\right),
\end{align}
which in the limit $\Lambda\to\infty$ reduce to be
\begin{align}
\Gamma_{xx}^{\mathrm{D}}(\mu,t,\Lambda)
&=8\mu\left[\frac{2}{\pi}\sqrt{\frac{2\Lambda}{\mu}}-1\right],\\
\Gamma_{yy}^{\mathrm{D}}(\mu,t,\Lambda)
&=8\mu,
\end{align}
indicating that $\Gamma_{xx}^{\mathrm{D}}(\mu,t,\Lambda)$ is divergent but $\Gamma_{yy}^{\mathrm{D}}(\mu,t,\Lambda)$ is convergent. By further analyzing $\Gamma_{xx}^{\mathrm{D}}(\mu,t,\Lambda)$ and $\Gamma_{yy}^{\mathrm{D}}(\mu,t,\Lambda)$
around $t=1$, it can be seen that when the tilted Dirac band undergoes a Lifshitz transition from the type-I phase
to the type-II phase, the Drude contribution $\Gamma_{yy}^{\mathrm{D}}(\mu,t,\Lambda)$ is continuous, whereas $\Gamma_{xx}^{\mathrm{D}}(\mu,t,\Lambda)$ changes from convergent to divergent.

For convenience, we plot the dependence of the real part of LOCs on the tilt parameter $t$ in Fig. \ref{fig4} after
introducing the relation
\begin{align}
\frac{\mathrm{Re}\sigma_{jj}(\omega)}{\sigma_0}&=\frac{v_x}{v_y}\Phi_{xx}(\omega)\delta_{jx} +\frac{v_y}{v_x}\Phi_{yy}(\omega)\delta_{jy},
\label{Eq57}
\end{align}
with
\begin{align}
\Phi_{xx}(\omega)&=\Gamma_{xx}^{\mathrm{IB}}(\omega)+\Gamma_{xx}^{\mathrm{D}}(\mu,t,\Lambda)\delta(\omega),
\label{Eq58}\\
\Phi_{yy}(\omega)&=\Gamma_{yy}^{\mathrm{IB}}(\omega)+\Gamma_{yy}^{\mathrm{D}}(\mu,t,\Lambda)\delta(\omega),
\label{Eq59}
\end{align}
and utilizing the analytical expressions listed in Secs. \ref{Sec:Interband conductivity} and  \ref{Sec:Drude conductivity}.
For numerical evaluation, we replace the Dirac $\delta$ function in the Drude conductivity with Lorentzians according to $\delta(x)\to(\eta/\pi)/(x^2+\eta^2)$ with $\eta\to0^{+}$. Note that all of the analytical results are independently confirmed by the direct numerical evaluation.

%%%%%%%%%%%%%%%%%%%%%%%%%%%%%%%%
\section{Summary and conclusions\label{Sec:CONCLUSIONS}}
%%%%%%%%%%%%%%%%%%%%%%%%%%%%%%%%

In this work, we theoretically investigated the LOCs in the type-I, type-II, and type-III phases of 2D tilted Dirac
energy bands. For the undoped case, the interband LOCs are constants either independent of the tilt parameter in both
type-I and type-III phases, or determined by the tilt parameter in the type-II phase. For the doped type-I or type-II
phase, the interband LOCs are anisotropic and share two critical boundaries at $\omega=\omega_1(t)$ and $\omega=\omega_2(t)$ 
confirmed by the JDOS. The tilt parameter $t$ and chemical potential $\mu$ can be extracted from optical experiments by 
measuring the positions of these two peaks and their separation $\Delta\omega(t)=\omega_2(t)-\omega_1(t)$. With increasing
the tilt parameter $t$ this separation becomes larger in the type-I phase whereas smaller in the type-II phase. At large
photon energy regime, the interband LOCs decay to certain asymptotic values which are exactly the same as that in the
undoped case. The Drude conductivity is also anisotropic and sensitive to the structure of the Fermi surface. The relation 
$\Gamma_{xx}^{\mathrm{D}}(\mu,t,\Lambda)/\Gamma_{yy}^{\mathrm{D}}(\mu,t,\Lambda)=1/\sqrt{1-t^2}$ always holds for the type-I 
phase. In the type-II phase, the Drude conductivity is closely related to the momentum cutoff $\Lambda$. When the 2D tilted 
Dirac band undergoes a Lifshitz transition, $\Gamma_{yy}^{\mathrm{D}}(\mu,t,\Lambda)$ is always convergent and continuous, 
but $\Gamma_{xx}^{\mathrm{D}}(\mu,t,\Lambda)$ change from convergent in the type-I phase to divergent in the type-II phase.

Through the shapes, asymptotic background values, critical boundaries and their separation in the optical conductivity
spectrum, such as obtained from the transmissivity and reflectivity measured by optical spectroscopy \cite{PRLMarel2008,
PRLMak2008,NatLi2008,PRBStauber2008}, the type of tilted Dirac bands can be determined in very clean samples at extremely
low temperature. These quantities can hence be taken as experimental signatures of the Lifshitz transition in the 2D
tilted Dirac materials. It is emphasized that the asymptotic background values of LOC do not depend on the chemical
potential, temperature, disorder and band gap \cite{PRBMoS22021,PRBCarbotte2016,PRLMikhailov2007,PRLMarel2008,
PRLMak2008,PRBStille2012,PRBAshby2014}. This can be physically understood by noticing that at large photon energy
regime the chemical potential, temperature, disorder, and band gap can be considered to be overwhelmingly small compared
to the photon energy $\omega$ or the energy $\varepsilon_\kappa^\lambda(k_x,k_y)$, and hence can be safely neglected in
analyzing the asymptotic behavior of LOCs. Consequently, the asymptotic values of LOC are robust against thermal
broadening or disorder, although the sharp critical boundaries in the LOCs may not survive. Hence, the physics of
Lifshitz transition can still be experimentally realized even in the presence of impurity and/or thermal broadening.

From a comprehensive comparison of LOCs \cite{PRLCarbotte2006,PRLMikhailov2007,PRLMarel2008,PRLMak2008,PRBStauber2008,
PRBStille2012,JPSJNishine2010, PRBMojarro2022,PRBVerma2017,PRBHerrera2019,PRBMoS22021,PRBJDOS2021}, it can be concluded
that the parameters of specific 2D Dirac materials, such as different anisotropic Fermi velocities and band gaps, do not
qualitatively affect the essential physical behaviors. Due to the underlying intrinsic similarities of 2D tilted Dirac
bands, the results of this work are expected to be qualitatively valid for a great number of 2D tilted Dirac materials, including $\alpha$-(BEDT-TTF)$_2$I$_3$, 8-$Pmmn$ borophene, $\alpha$-SnS$_2$, TaCoTe$_2$, TaIrTe$_4$, and different compounds of $1T^\prime$ transition metal dichalcogenides.

\emph{Note added.}  Recently, a related paper appeared \cite{PRBWild2022}, which also examines similar topics,
such as, the effects of different tilting and anisotropic Fermi velocity on the optical response in 2D tilted Dirac
materials and confirms partial findings discussed herein.

%%%%%%%%%%%%%%%%%%%%%%%%%%%%%%%
\section*{ACKNOWLEDGEMENTS\label{Sec:acknowledgements}}
%%%%%%%%%%%%%%%%%%%%%%%%%%%%%%%

We are grateful to Xiang-Hua Kong, Zhi-Qiang Li, and Chen-Yi Zhou for valuable discussions. This work is
partially supported by the National Natural Science Foundation of China (NNSFC) under Grant No. 11547200. C.-Y.T. and
J.-T.H. acknowledges financial support from the NNSFC under Grants No.11874273 and No. 11874271, respectively. H.-R.C.
and H.G. are also supported by the NSERC of Canada and the FQRNT of Quebec (H.G.). We thank the High Performance
Computing Centers at Sichuan Normal University,  McGill University, and Compute Canada. H.-R.C. would like to dedicate
this paper to the memory of his past mother Xiu-Lan Xu.

\appendix

\begin{widetext}

\section{Definition and particle-hole symmetry of LOC and JDOS\label{Sec:AppendixA}}
In this appendix, we give the definition of LOC and JDOS, and prove their particle-hole symmetry.

\subsection{Definition and Particle-hole symmetry of LOC}
Within linear response theory, the LOC for the photon frequency $\omega$ and chemical potential $\mu$ is given by

\begin{align}
\sigma_{jj}(\omega,\mu)&=g_s\sum_{\kappa=\pm}\sigma_{jj}^\kappa(\omega,\mu),
\end{align}
where
\begin{align}
\sigma_{jj}^\kappa(\omega,\mu)&= \frac{i}{\omega} \int_{-\infty}^{+\infty}\frac{dk_x}{2\pi} \int_{-\infty}^{+\infty}\frac{dk_y}{2\pi}\sum_{\lambda=\pm}\sum_{\lambda^\prime=\pm} \mathcal{F}_{\lambda,\lambda^\prime}^{\kappa;jj}(k_x,k_y)\frac{f\left[\varepsilon_\kappa^{\lambda}(k_x,k_y),\mu\right] -f\left[\varepsilon_\kappa^{\lambda^\prime}(k_x,k_y),\mu\right]}{\omega+\varepsilon_\kappa^{\lambda}(k_x,k_y)- \varepsilon_\kappa^{\lambda^\prime}(k_x,k_y)+i\eta}.
\label{A5}
\end{align}
Here we keep the explicit dependence of $\mu$ temporarily for the sake of proving the particle-hole symmetry. In these two equations, $j=x,y$ refer to spatial coordinates, the conduction band and valence band are denoted respectively by $\lambda=+$ and $\lambda=-$, $\eta$ denotes a positive infinitesimal, and $f(x,\mu)$ is the Fermi distribution function. In addition, $f(x,\mu)$, $\varepsilon_\kappa^\lambda(k_x,k_y)$ and $\mathcal{F}_{\lambda,\lambda^\prime}^{\kappa;jj}(k_x,k_y)$ are given as
\begin{align}
f(x,\mu)&=\frac{1}{1+\exp\left[\beta(x-\mu)\right]},\\
\varepsilon_\kappa^\lambda(k_x,k_y)&=\kappa t v_y k_y+\lambda \mathcal{Z}(k_x,k_y),\\
\mathcal{F}_{\lambda,\lambda^\prime}^{\kappa;xx}(k_x,k_y)
=&\frac{e^2}{2}v_x^2 \left\{1+\lambda\lambda^\prime\frac{v_x^2k_x^2-v_y^2k_y^2} {\left[\mathcal{Z}(k_x,k_y)\right]^2}\right\},\\
\mathcal{F}_{\lambda,\lambda^\prime}^{\kappa;yy} (k_x,k_y)
=&\frac{e^2}{2}v_y^2\left\{t^2(1+\lambda\lambda^\prime)+
1+\lambda\lambda^\prime\frac{-v_x^2k_x^2+v_y^2k_y^2}{\left[\mathcal{Z}(k_x,k_y)\right]^2}
+2(\lambda+\lambda^\prime)\frac{\kappa t v_y k_y}{\mathcal{Z}(k_x,k_y)}
\right\}
\nonumber\\
=&\frac{e^2}{2}v_y^2\left\{2t^2\delta_{\lambda\lambda^\prime}+
1-\lambda\lambda^\prime\frac{v_x^2k_x^2-v_y^2k_y^2}{\left[\mathcal{Z}(k_x,k_y)\right]^2}
+4\lambda\delta_{\lambda\lambda^\prime}\frac{\kappa t v_y k_y}{\mathcal{Z}(k_x,k_y)}
\right\},
\end{align}
where $\beta=1/k_BT$ and $\mathcal{Z}(k_x,k_y)=\sqrt{v_x^2 k_x^2+v_y^2 k_y^2}$. It can be verified that
\begin{align}
f(x,\mu)&=1-f(-x,-\mu),\\
\varepsilon_\kappa^\lambda(k_x,k_y)&=-\varepsilon_{-\kappa}^{-\lambda}(k_x,k_y),\\
\mathcal{F}_{\lambda,\lambda^\prime}^{\kappa;jj}(k_x,k_y)&=\mathcal{F}_{-\lambda,-\lambda^\prime}^{-\kappa;jj}(k_x,k_y)
=\mathcal{F}_{\lambda^\prime,\lambda}^{\kappa;jj}(k_x,k_y).
\end{align}

By utilizing these relations, we have
\begin{align}
&\sigma_{jj}(\omega,\mu)=g_s\frac{i}{\omega} \int_{-\infty}^{+\infty}\frac{dk_x}{2\pi} \int_{-\infty}^{+\infty}\frac{dk_y}{2\pi}\sum_{\kappa=\pm}\sum_{\lambda=\pm}\sum_{\lambda^\prime=\pm} \mathcal{F}_{\lambda,\lambda^\prime}^{\kappa;jj}(k_x,k_y)
\frac{f\left[\varepsilon_\kappa^{\lambda}(k_x,k_y),\mu\right] -f\left[\varepsilon_\kappa^{\lambda^\prime}(k_x,k_y),\mu\right]}{\omega
+\varepsilon_\kappa^{\lambda}(k_x,k_y)-\varepsilon_\kappa^{\lambda^\prime}(k_x,k_y)+i\eta}\notag\\
&=g_s\frac{i}{\omega} \int_{-\infty}^{+\infty}\frac{dk_x}{2\pi} \int_{-\infty}^{+\infty}\frac{dk_y}{2\pi}\sum_{\kappa=\pm}\sum_{\lambda=\pm}\sum_{\lambda^\prime=\pm} \mathcal{F}_{\lambda,\lambda^\prime}^{\kappa;jj}(k_x,k_y)
\frac{f\left[-\varepsilon_\kappa^{\lambda^\prime}(k_x,k_y),-\mu\right]
-f\left[-\varepsilon_\kappa^{\lambda}(k_x,k_y),-\mu\right]}{\omega
+\left[-\varepsilon_\kappa^{\lambda^\prime}(k_x,k_y)\right]
-\left[-\varepsilon_\kappa^{\lambda}(k_x,k_y)\right]+i\eta}\notag\\
&=g_s\frac{i}{\omega} \int_{-\infty}^{+\infty}\frac{dk_x}{2\pi} \int_{-\infty}^{+\infty}\frac{dk_y}{2\pi}\sum_{\kappa=\pm}\sum_{\lambda=\pm}\sum_{\lambda^\prime=\pm} \mathcal{F}_{-\lambda^\prime,-\lambda}^{-\kappa;jj}(k_x,k_y)
\frac{f\left[-\varepsilon_{-\kappa}^{-\lambda}(k_x,k_y),-\mu\right]
-f\left[\varepsilon_{-\kappa}^{-\lambda^\prime}(k_x,k_y),-\mu\right]}{\omega
+\left[-\varepsilon_{-\kappa}^{-\lambda}(k_x,k_y)\right]
-\left[\varepsilon_{-\kappa}^{-\lambda^\prime}(k_x,k_y)\right]+i\eta}\notag\\
&=g_s\frac{i}{\omega} \int_{-\infty}^{+\infty}\frac{dk_x}{2\pi} \int_{-\infty}^{+\infty}\frac{dk_y}{2\pi}\sum_{\kappa=\pm}\sum_{\lambda=\pm}\sum_{\lambda^\prime=\pm} \mathcal{F}_{\lambda^\prime,\lambda}^{\kappa;jj}(k_x,k_y)
\frac{f\left[\varepsilon_{\kappa}^{\lambda}(k_x,k_y),-\mu\right]
-f\left[\varepsilon_{\kappa}^{\lambda^\prime}(k_x,k_y),-\mu\right]}{\omega+\varepsilon_{\kappa}^{\lambda}(k_x,k_y)
-\varepsilon_{\kappa}^{\lambda^\prime}(k_x,k_y)+i\eta}\notag\\
&=g_s\frac{i}{\omega} \int_{-\infty}^{+\infty}\frac{dk_x}{2\pi} \int_{-\infty}^{+\infty}\frac{dk_y}{2\pi}\sum_{\kappa=\pm}\sum_{\lambda=\pm}\sum_{\lambda^\prime=\pm} \mathcal{F}_{\lambda,\lambda^\prime}^{\kappa;jj}(k_x,k_y)
\frac{f\left[\varepsilon_{\kappa}^{\lambda}(k_x,k_y),-\mu\right]
-f\left[\varepsilon_{\kappa}^{\lambda^\prime}(k_x,k_y),-\mu\right]}{\omega+\varepsilon_{\kappa}^{\lambda}(k_x,k_y)
-\varepsilon_{\kappa}^{\lambda^\prime}(k_x,k_y)+i\eta}\notag\\
&=\sigma_{jj}(\omega,-\mu),
\end{align}
which indicates $\sigma_{jj}(\omega,\mu)$ respects the particle-hole symmetry, namely,
\begin{align}
\sigma_{jj}(\omega,\mu)&=\sigma_{jj}(\omega,-\mu)=\sigma_{jj}(\omega,|\mu|).
\end{align}

Keeping this property of $\sigma_{jj}(\omega,\mu)$ in mind, we can safely replace $\mu$ in all of $\sigma_{jj}(\omega,\mu)$, $\sigma_{jj}^{\kappa}(\omega,\mu)$, and $f(x,\mu)$ by $|\mu|$ since we are only concerned with the final result of $\sigma_{jj}(\omega,\mu)$. Hereafter, we restrict our analysis to the $n$-doped case ($\mu>0$). For simplicity, we further denote $\sigma_{jj}(\omega,\mu )\equiv\sigma_{jj}(\omega)$, $\sigma_{jj}^\kappa(\omega,\mu )\equiv\sigma_{jj}^\kappa(\omega)$, and $f(x,\mu )\equiv f(x)$. Consequently, we have
\begin{align}
\sigma_{jj}(\omega)&\equiv\sigma_{jj}(\omega,\mu )= g_s\frac{i}{\omega} \int_{-\infty}^{+\infty}\frac{dk_x}{2\pi} \int_{-\infty}^{+\infty}\frac{dk_y}{2\pi}\sum_{\kappa=\pm}\sum_{\lambda=\pm}\sum_{\lambda^\prime=\pm} \mathcal{F}_{\lambda,\lambda^\prime}^{\kappa;jj}(k_x,k_y)\frac{f\left[\varepsilon_\kappa^{\lambda}(k_x,k_y),\mu \right] -f\left[\varepsilon_\kappa^{\lambda^\prime}(k_x,k_y),\mu \right]}{\omega+\varepsilon_\kappa^{\lambda}(k_x,k_y)- \varepsilon_\kappa^{\lambda^\prime}(k_x,k_y)+i\eta}\notag\\
&\equiv g_s \frac{i}{\omega} \int_{-\infty}^{+\infty}\frac{dk_x}{2\pi} \int_{-\infty}^{+\infty}\frac{dk_y}{2\pi}\sum_{\kappa=\pm}\sum_{\lambda=\pm}\sum_{\lambda^\prime=\pm} \mathcal{F}_{\lambda,\lambda^\prime}^{\kappa;jj}(k_x,k_y)\frac{f\left[\varepsilon_\kappa^{\lambda}(k_x,k_y)\right] -f\left[\varepsilon_\kappa^{\lambda^\prime}(k_x,k_y)\right]}{\omega+\varepsilon_\kappa^{\lambda}(k_x,k_y)- \varepsilon_\kappa^{\lambda^\prime}(k_x,k_y)+i\eta},\\
\sigma_{jj}^\kappa(\omega)&\equiv\sigma_{jj}^\kappa(\omega,\mu )
= \frac{i}{\omega} \int_{-\infty}^{+\infty}\frac{dk_x}{2\pi} \int_{-\infty}^{+\infty}\frac{dk_y}{2\pi}\sum_{\lambda=\pm}\sum_{\lambda^\prime=\pm}
\mathcal{F}_{\lambda,\lambda^\prime}^{\kappa;jj}(k_x,k_y)\frac{f\left[\varepsilon_\kappa^{\lambda}(k_x,k_y),\mu \right] -f\left[\varepsilon_\kappa^{\lambda^\prime}(k_x,k_y),\mu \right]}{\omega+\varepsilon_\kappa^{\lambda}(k_x,k_y)- \varepsilon_\kappa^{\lambda^\prime}(k_x,k_y)+i\eta}\notag\\
&\equiv \frac{i}{\omega} \int_{-\infty}^{+\infty}\frac{dk_x}{2\pi} \int_{-\infty}^{+\infty}\frac{dk_y}{2\pi}\sum_{\lambda=\pm}\sum_{\lambda^\prime=\pm}
\mathcal{F}_{\lambda,\lambda^\prime}^{\kappa;jj}(k_x,k_y)\frac{f\left[\varepsilon_\kappa^{\lambda}(k_x,k_y)\right] -f\left[\varepsilon_\kappa^{\lambda^\prime}(k_x,k_y)\right]}{\omega+\varepsilon_\kappa^{\lambda}(k_x,k_y)- \varepsilon_\kappa^{\lambda^\prime}(k_x,k_y)+i\eta}.
\end{align}

\subsection{Definition and particle-hole symmetry of JDOS}

The JDOS for the photon frequency $\omega$ and chemical potential $\mu$ is given by
\begin{align}
\mathcal{J}(\omega,\mu)=g_s \sum_{\kappa=\pm}\mathcal{J}_\kappa(\omega,\mu),
\end{align}
where
\begin{align}
\mathcal{J}_{\kappa}(\omega,\mu)
&=\int\frac{d^2\boldsymbol{k}}{(2\pi)^2}\delta\left[\varepsilon_\kappa^+(k_x,k_y)-\varepsilon_\kappa^-(k_x,k_y)-\omega\right] \left\{\Theta\left[\mu-\varepsilon_\kappa^-(k_x,k_y)\right]- \Theta\left[\mu-\varepsilon_\kappa^+(k_x,k_y)\right]\right\}.
\end{align}

Here we keep the explicit dependence of $\mu$ temporarily for the sake of proving the particle-hole symmetry.

After introducing $\tilde{k}_x=v_xk_x$ and $\tilde{k}_y=v_yk_y$, the JDOS can be written as
\begin{align}
\mathcal{J}_\kappa(\omega,\mu)
&= \frac{1}{4\pi^2v_xv_y} \int d\tilde{k}_xd\tilde{k}_y
\delta\left[\tilde{\varepsilon}_\kappa^+(\tilde{k}_x,\tilde{k}_y)-\tilde{\varepsilon}_\kappa^-(\tilde{k}_x,\tilde{k}_y)-\omega\right]
\left\{\Theta\left[\mu-\tilde{\varepsilon}_\kappa^-(\tilde{k}_x,\tilde{k}_y)\right]- \Theta\left[\mu-\tilde{\varepsilon}_\kappa^+(\tilde{k}_x,\tilde{k}_y)\right]\right\}
\notag\\&
= \frac{\mathcal{J}_0}{2\pi} \int d\tilde{k}_xd\tilde{k}_y
\delta\left[\tilde{\varepsilon}_\kappa^+(\tilde{k}_x,\tilde{k}_y)-\tilde{\varepsilon}_\kappa^-(\tilde{k}_x,\tilde{k}_y)-\omega\right]
\left\{\Theta\left[\mu-\tilde{\varepsilon}_\kappa^-(\tilde{k}_x,\tilde{k}_y)\right]- \Theta\left[\mu-\tilde{\varepsilon}_\kappa^+(\tilde{k}_x,\tilde{k}_y)\right]\right\},
\end{align}
where $\mathcal{J}_0=1/(2\pi v_xv_y)$ and $\tilde{\varepsilon}_\kappa^{\lambda}(\tilde{k}_x,\tilde{k}_y)=\kappa t \tilde{k}_y +\lambda\sqrt{\tilde{k}_x^2+\tilde{k}_y^2}$.

In the polar coordinate, $\tilde{\varepsilon}_\kappa^{\lambda}(\tilde{k}_x,\tilde{k}_y)=\left[\lambda+\kappa t \sin\phi \right]\tilde{k}$ and $\tilde{\varepsilon}_\kappa^+(\tilde{k}_x,\tilde{k}_y)-\tilde{\varepsilon}_\kappa^-(\tilde{k}_x,\tilde{k}_y) =2\tilde{k}$. Solving the equation
\begin{align}
\tilde{\varepsilon}_\kappa^{\lambda}(\tilde{k}_x,\tilde{k}_y)&=\left[\lambda+\kappa t \sin\phi \right]
\tilde{k}_F(\kappa,\lambda,\mu)=\mu,
\end{align}
we have the Fermi wave vectors
\begin{align}
\tilde{k}_F(\kappa,\lambda,\mu)
&=\frac{\mu}{\lambda+\kappa t \sin\phi}
=\frac{\mathrm{sgn}(\mu)|\mu|}{\lambda+\kappa t \sin\phi}
=\frac{|\mu|}{\mathrm{sgn}(\mu)\lambda+\mathrm{sgn}(\mu)\kappa t \sin\phi}
\notag\\&
=\tilde{k}_F^{\kappa,\lambda}(\mu)\left[\Theta(1-t)\delta_{\lambda,\mathrm{sgn}(\mu)}
+\Theta(t-1)\left(\delta_{\lambda,\mathrm{sgn}(\mu)}+\delta_{\lambda,-\mathrm{sgn}(\mu)}\right)\right]
\end{align}
satisfying the relation
\begin{align}
\tilde{k}_F^{\kappa,\lambda}(-\mu)&=\tilde{k}_F^{-\kappa,-\lambda}(\mu).
\end{align}

From the above two relations, the JDOS can be rewritten as
\begin{align}
\mathcal{J}(\omega,\mu)&=
\tilde{\Theta}(t)\left[\tilde{\Theta}(1-t)+\Theta(t-1)\right]\mathcal{J}(\omega,\mu)
=\tilde{\Theta}(t)\left[\tilde{\Theta}(1-t)+\Theta(t-1)\right]\sum_{\kappa=\pm}\mathcal{J}_\kappa(\omega,\mu)
\notag\\&
=\tilde{\Theta}(t)\left[\tilde{\Theta}(1-t)+\Theta(t-1)\right]\sum_{\kappa=\pm}
\frac{\mathcal{J}_0}{2\pi}\int d\tilde{k}_xd\tilde{k}_y\delta\left[2\tilde{k}-\omega\right] \left\{\Theta\left[\mu-\tilde{\varepsilon}_\kappa^-(\tilde{k}_x,\tilde{k}_y)\right]- \Theta\left[\mu-\tilde{\varepsilon}_\kappa^+(\tilde{k}_x,\tilde{k}_y)\right]\right\}
\notag\\&
=\tilde{\Theta}(t)\tilde{\Theta}(1-t)\frac{\mathcal{J}_0}{2\pi}
\int_{-\infty}^{\infty}\int_{-\infty}^{\infty}d\tilde{k}_xd\tilde{k}_y
\delta\left[2\tilde{k}-\omega\right]
\sum_{\kappa=\pm}\sum_{\lambda=\pm}
\Theta\left[\omega-2\tilde{k}_F(\kappa,\lambda,\mu)\right]
\notag\\&
+\tilde{\Theta}(t)\Theta(t-1)\frac{\mathcal{J}_0}{2\pi}
\int_{-\infty}^{\infty}\int_{-\infty}^{\infty}d\tilde{k}_xd\tilde{k}_y
\delta\left[2\tilde{k}-\omega\right]
\sum_{\kappa=\pm}\sum_{\lambda=\pm}
\Theta\left\{\mathrm{sgn}(\mu)\left[\tilde{k}_F^{\kappa,-}(\mu)-\tilde{k}_F^{\kappa,+}(\mu)
\right]\right\}
\Theta\left[\omega-2\tilde{k}_F(\kappa,\lambda,\mu)\right]
\notag\\&
=\tilde{\Theta}(t)\tilde{\Theta}(1-t)\frac{\mathcal{J}_0}{2\pi}
\int_{-\infty}^{\infty}\int_{-\infty}^{\infty}d\tilde{k}_xd\tilde{k}_y
\delta\left[2\tilde{k}-\omega\right]
\sum_{\kappa=\pm}\sum_{\lambda=\pm}
\Theta\left[\omega-2\tilde{k}_F^{\kappa,\lambda}(\mu)\delta_{\lambda,\mathrm{sgn}(\mu)}\right].
\notag\\&
+\tilde{\Theta}(t)\Theta(t-1)\frac{\mathcal{J}_0}{2\pi}
\int_{-\infty}^{\infty}\int_{-\infty}^{\infty}d\tilde{k}_xd\tilde{k}_y
\delta\left[2\tilde{k}-\omega\right]
\sum_{\kappa=\pm}\sum_{\lambda=\pm}
\Theta\left\{\mathrm{sgn}(\mu)\left[\tilde{k}_F^{\kappa,-}(\mu)-\tilde{k}_F^{\kappa,+}(\mu)
\right]\right\}
\notag\\&\times
\left\{\Theta\left[\omega-2\tilde{k}_F^{\kappa,\lambda}(\mu)\delta_{\lambda,\mathrm{sgn}(\mu)}\right]-
\Theta\left[\omega-2\tilde{k}_F^{\kappa,\lambda}(\mu)\delta_{\lambda,\mathrm{sgn}(-\mu)}\right]\right\}.
\end{align}
Consequently, we have
\begin{align}
\mathcal{J}(\omega,-\mu)&=
\tilde{\Theta}(t)\tilde{\Theta}(1-t)\frac{\mathcal{J}_0}{2\pi}
\int_{-\infty}^{\infty}\int_{-\infty}^{\infty}d\tilde{k}_xd\tilde{k}_y
\delta\left[2\tilde{k}-\omega\right]
\sum_{\kappa=\pm}\sum_{\lambda=\pm}
\Theta\left[\omega-2\tilde{k}_F^{\kappa,\lambda}(-\mu)\delta_{\lambda,\mathrm{sgn}(-\mu)}\right]
\notag\\&
+\tilde{\Theta}(t)\Theta(t-1)\frac{\mathcal{J}_0}{2\pi}
\int_{-\infty}^{\infty}\int_{-\infty}^{\infty}d\tilde{k}_xd\tilde{k}_y
\delta\left[2\tilde{k}-\omega\right]
\sum_{\kappa=\pm}\sum_{\lambda=\pm}
\Theta\left\{\mathrm{sgn}(-\mu)\left[\tilde{k}_F^{\kappa,-}(-\mu)-\tilde{k}_F^{\kappa,+}(-\mu)
\right]\right\}
\notag\\&\times
\left\{\Theta\left[\omega-2\tilde{k}_F^{\kappa,\lambda}(-\mu)\delta_{\lambda,\mathrm{sgn}(-\mu)}\right]-
\Theta\left[\omega-2\tilde{k}_F^{\kappa,\lambda}(-\mu)\delta_{\lambda,-\mathrm{sgn}(-\mu)}\right]\right\}
\notag\\&
=\tilde{\Theta}(t)\tilde{\Theta}(1-t)\frac{\mathcal{J}_0}{2\pi}
\int_{-\infty}^{\infty}\int_{-\infty}^{\infty}d\tilde{k}_xd\tilde{k}_y
\delta\left[2\tilde{k}-\omega\right]
\sum_{\kappa=\pm}\sum_{\lambda=\pm}
\Theta\left[\omega-2\tilde{k}_F^{-\kappa,-\lambda}(\mu)\delta_{-\lambda,\mathrm{sgn}(\mu)}\right]
\notag\\&
+\tilde{\Theta}(t)\Theta(t-1)\frac{\mathcal{J}_0}{2\pi}
\int_{-\infty}^{\infty}\int_{-\infty}^{\infty}d\tilde{k}_xd\tilde{k}_y
\delta\left[2\tilde{k}-\omega\right]
\sum_{\kappa=\pm}\sum_{\lambda=\pm}
\Theta\left\{-\mathrm{sgn}(\mu)\left[\tilde{k}_F^{-\kappa,+}(\mu)-\tilde{k}_F^{-\kappa,-}(\mu)
\right]\right\}
\notag\\&\times
\left\{\Theta\left[\omega-2\tilde{k}_F^{-\kappa,-\lambda}(\mu)\delta_{-\lambda,\mathrm{sgn}(\mu)}\right]-
\Theta\left[\omega-2\tilde{k}_F^{-\kappa,-\lambda}(\mu)\delta_{-\lambda,\mathrm{sgn}(-\mu)}\right]\right\}
\notag\\&
=\tilde{\Theta}(t)\tilde{\Theta}(1-t)\frac{\mathcal{J}_0}{2\pi}
\int_{-\infty}^{\infty}\int_{-\infty}^{\infty}d\tilde{k}_xd\tilde{k}_y
\delta\left[2\tilde{k}-\omega\right]
\sum_{\kappa=\pm}\sum_{\lambda=\pm}
\Theta\left[\omega-2\tilde{k}_F^{\kappa,\lambda}(\mu)\delta_{\lambda,\mathrm{sgn}(\mu)}\right]
\notag\\&
+\tilde{\Theta}(t)\Theta(t-1)\frac{\mathcal{J}_0}{2\pi}
\int_{-\infty}^{\infty}\int_{-\infty}^{\infty}d\tilde{k}_xd\tilde{k}_y
\delta\left[2\tilde{k}-\omega\right]
\sum_{\kappa=\pm}\sum_{\lambda=\pm}
\Theta\left\{\mathrm{sgn}(\mu)\left[\tilde{k}_F^{\kappa,-}(\mu)-\tilde{k}_F^{\kappa,+}(\mu)
\right]\right\}
\notag\\&\times
\left\{\Theta\left[\omega-2\tilde{k}_F^{\kappa,\lambda}(\mu)\delta_{\lambda,\mathrm{sgn}(\mu)}\right]-
\Theta\left[\omega-2\tilde{k}_F^{\kappa,\lambda}(\mu)\delta_{\lambda,\mathrm{sgn}(-\mu)}\right]\right\}
\notag\\&
=\mathcal{J}(\omega,\mu).
\end{align}

Obviously, $\mathcal{J}(\omega,\mu)$ respects the particle-hole symmetry, namely,
\begin{align}
\mathcal{J}(\omega,\mu)&=\mathcal{J}(\omega,-\mu)=\mathcal{J}(\omega,|\mu|).
\end{align}

Keeping this property of $\mathcal{J}(\omega,\mu)$ in mind, we can safely replace $\mu$ in all of $\mathcal{J}(\omega,\mu)$, $\mathcal{J}_{\kappa}(\omega,\mu)$, and $\Theta(x,\mu)$ by $|\mu|$ since we are only concerned with the final result of $\mathcal{J}(\omega,\mu)$. Hereafter, we are allowed to restrict our analysis to the n-doped case ($\mu>0$). For simplicity, we further denote $\mathcal{J}(\omega,\mu )\equiv\mathcal{J}(\omega)$, $\mathcal{J}_\kappa(\omega,\mu )\equiv\mathcal{J}_\kappa(\omega)$, and $\Theta(x,\mu )\equiv \Theta(x)$.

\section{Detailed calculation of LOC \label{Sec:AppendixB}}

In the following, we present the detailed calculation of them in the $n$-doped case ($\mu>0$). Before doing that, we factor the ratio $v_y/v_x$ or $v_x/v_y$ out from the original expressions in order to simplify the calculation. This way, the original expressions in the \emph{anisotropic model} ($v_x\neq v_y$ and $v_t\ge0$) are converted to be the rescaled forms in the \emph{isotropic model} ($v_x= v_y$ and $v_t\ge0$). Next we focus on the interband part and intraband part of LOC, respectively.

\subsection{Calculation of interband LOC}
After substituting $\tilde{k}_x=v_x k_x$ and $\tilde{k}_y=v_y k_y$ into Eq.(\ref{Eq5}) and expressing it in terms of $\tilde{k}=\sqrt{\tilde{k}_x^2+\tilde{k}_y^2}$ and $\phi=\arctan(\tilde{k}_y/\tilde{k}_x)$, the original expressions in the anisotropic model are converted to be the rescaled forms in the isotropic model as
\begin{align}
\mathrm{Re}\sigma_{jj(\mathrm{IB})}^\kappa(\omega)
=\int \frac{d\tilde{k}_xd\tilde{k}_y}{4\pi v_x v_y}\mathcal{\tilde{F}}_{-,+}^{\kappa;jj}(\tilde{k}_x,\tilde{k}_y) \frac{f\left[\kappa t \tilde{k}_y-\tilde{k}\right]-f\left[\kappa t\tilde{k}_y+\tilde{k}\right]}{\omega}\delta\left[\omega-2\tilde{k}\right],
\end{align}
where
\begin{align}
\mathcal{\tilde{F}}_{-,+}^{\kappa;xx}(\tilde{k}_x,\tilde{k}_y)&=4\sigma_0\frac{v_x^2\tilde{k}_y^2}{\tilde{k}^2} =4\sigma_0v_x^2\sin^2\phi,\notag\\
\mathcal{\tilde{F}}_{-,+}^{\kappa;yy}(\tilde{k}_x,\tilde{k}_y)&=4\sigma_0\frac{v_y^2\tilde{k}_x^2}{\tilde{k}^2} =4\sigma_0v_y^2\cos^2\phi.\notag
\end{align}

As a consequence, the ratio $v_y/v_x$ or $v_x/v_y$ can be factored out from the original expressions as
\begin{align}
\mathrm{Re}\sigma_{xx(\mathrm{IB})}^\kappa(\omega)
&=\sigma_0\frac{v_x}{v_y}\Gamma_{xx(\mathrm{IB})}^{\kappa}(\omega),\\
\mathrm{Re}\sigma_{yy(\mathrm{IB})}^\kappa(\omega)
&=\sigma_0\frac{v_y}{v_x}\Gamma_{yy(\mathrm{IB})}^{\kappa}(\omega),
\end{align}
where two dimensionless auxiliary functions
\begin{align}
\Gamma_{xx(\mathrm{IB})}^{\kappa}(\omega)
&=\int_0^{+\infty}\frac{\tilde{k}d\tilde{k}}{\omega} \int_{0}^{2\pi}\frac{\sin^2\phi d\phi}{\pi}\delta\left[\omega-2\tilde{k}\right]
\left\{f\left[(\kappa t \sin\phi-1)\tilde{k}\right]-f\left[(\kappa t \sin\phi+1)\tilde{k}\right] \right\}\notag\\
&=\int_0^{+\infty}\frac{\tilde{k}d\tilde{k}}{\omega} \int_{-\pi/2}^{3\pi/2}\frac{\sin^2\phi d\phi}{\pi}\delta\left[\omega-2\tilde{k}\right]
\left\{f\left[(\kappa t \sin\phi-1)\tilde{k}\right]-f\left[(\kappa t \sin\phi+1)\tilde{k}\right] \right\},\\
%%%%%%%%%%%%%%%%%%%%%%%%%%%%%%%%%%%%%%%%%%%
\Gamma_{yy(\mathrm{IB})}^{\kappa}(\omega)
&=\int_0^{+\infty}\frac{\tilde{k}d\tilde{k}}{\omega} \int_{-\pi/2}^{3\pi/2}\frac{\cos^2\phi d\phi}{\pi}\delta\left[\omega-2\tilde{k}\right]
\left\{f\left[(\kappa t \sin\phi-1)\tilde{k}\right]-f\left[(\kappa t \sin\phi+1)\tilde{k}\right] \right\},
\end{align}
are introduced for convenience. The ratio $v_x/v_y$ in $\mathrm{Re}\sigma_{xx(\mathrm{IB})}^\kappa(\omega)$ and $v_y/v_x$ in $\mathrm{Re}\sigma_{yy(\mathrm{IB})}^\kappa(\omega)$ are totally different in the anisotropic model but the same in the isotropic model. Therefore, the calculation of $\mathrm{Re}\sigma_{xx(\mathrm{IB})}^\kappa(\omega)$ and $\mathrm{Re}\sigma_{yy(\mathrm{IB})}^\kappa(\omega)$ boils down to calculating $\Gamma_{xx(\mathrm{IB})}^{\kappa}(\omega)$ and $\Gamma_{yy(\mathrm{IB})}^{\kappa}(\omega)$. Integrating over $\tilde{k}$ leads us to
\begin{align}
&\Gamma_{xx(\mathrm{IB})}^{\kappa}(\omega)=\frac{1}{4} \int_{-\pi/2}^{3\pi/2} \frac{\sin^2\phi d\phi}{\pi}
\left\{f\left[(\kappa t \sin\phi-1)\frac{\omega}{2}\right]
-f\left[(\kappa t \sin\phi+1)\frac{\omega}{2}\right]
\right\},\notag\\
&=\frac{1}{4}\int_{-\pi/2}^{\pi/2}\frac{\sin^2\phi d\phi}{\pi}
\left\{
f\left[(\kappa t \sin\phi-1)\frac{\omega}{2}\right]
-f\left[(\kappa t \sin\phi+1)\frac{\omega}{2}\right]
+f\left[(-\kappa t \sin\phi-1)\frac{\omega}{2}\right]
-f\left[(-\kappa t \sin\phi+1)\frac{\omega}{2}\right]\right\},\notag
\end{align}
and hence
\begin{align}
\Gamma_{xx}^{\mathrm{IB}}(\omega) &=g_s\left[\Gamma_{xx(\mathrm{IB})}^{+}(\omega)+\Gamma_{xx(\mathrm{IB})}^{-}(\omega)\right]\notag\\
&=\int_{-\pi/2}^{\pi/2}\frac{\sin^2\phi d\phi}{\pi}
\left\{f\left[(t \sin\phi-1)\frac{\omega}{2}\right]
-f\left[(t \sin\phi+1)\frac{\omega}{2}\right]
+f\left[(-t\sin\phi-1)\frac{\omega}{2}\right]-f\left[(-t \sin\phi+1)\frac{\omega}{2}\right]\right\},\notag
\end{align}
which includes the contribution of different valleys, where $g_s=2$ is the degeneracy parameter of spin. Parallel procedures give rise to
\begin{align}
&\Gamma_{yy}^{\mathrm{IB}}(\omega) =g_s\left[\Gamma_{yy(\mathrm{IB})}^{+}(\omega)+\Gamma_{yy(\mathrm{IB})}^{-}(\omega)\right]\notag\\
&=\int_{-\pi/2}^{\pi/2}\frac{\cos^2\phi d\phi}{\pi}
\left\{f\left[(t \sin\phi-1)\frac{\omega}{2}\right]
-f\left[(t \sin\phi+1)\frac{\omega}{2}\right]+f\left[(-t \sin\phi-1)\frac{\omega}{2}\right]-f\left[(-t \sin\phi+1)\frac{\omega}{2}\right]\right\}.
\end{align}

In order to obtain the analytical expressions, we perform the integrations over $\phi$ at zero temperature where
the Fermi distribution function $f(x)$ can be replaced by the Heaviside step function $\Theta[\mu -x]$ and consequently have
\begin{align}
\Gamma_{xx}^{\mathrm{IB}}(\omega) =&\int_{-\pi/2}^{\pi/2}\frac{2\sin^2\phi d\phi}{\pi}
\left\{\Theta\left[\mu -(t \sin\phi-1)\frac{\omega}{2}\right]-\Theta\left[\mu -(t \sin\phi+1)\frac{\omega}{2}\right]\right\}\notag\\
=&\int_{-1}^{1}\frac{2x^2 dx}{\pi\sqrt{1-x^2}}
\left\{\Theta\left[\mu -(t x-1)\frac{\omega}{2}\right]-\Theta\left[\mu -(t x+1)\frac{\omega}{2}\right]\right\},
\end{align}
and
\begin{align}
\Gamma_{yy}^{\mathrm{IB}}(\omega) =&\int_{-\pi/2}^{\pi/2}\frac{2\cos^2\phi d\phi}{\pi}
\left\{\Theta\left[\mu -(t \sin\phi-1)\frac{\omega}{2}\right]-\Theta\left[\mu -(t \sin\phi+1)\frac{\omega}{2}\right]\right\}\notag\\
=&\int_{-1}^{1}\frac{2\sqrt{1-x^2} dx}{\pi}
\left\{\Theta\left[\mu -(t x-1)\frac{\omega}{2}\right]-\Theta\left[\mu -(t x+1)\frac{\omega}{2}\right]\right\}.
\end{align}

After rewriting the Heaviside step function, we get
\begin{align}
\Gamma_{xx}^{\mathrm{IB}}(\omega)
=&\int_{-1}^{1}\frac{2x^2 dx}{\pi\sqrt{1-x^2}}
\left\{\Theta\left[-x+\frac{2\mu/\omega+1}{t}\right]
-\Theta\left[-x+\frac{2\mu/\omega-1}{t}\right]\right\}\notag\\
=&\int_{-1}^{1}\frac{2x^2 dx}{\pi\sqrt{1-x^2}}
\left\{\Theta\left[-x+\xi_{+}\right]-\Theta\left[-x+\xi_{-}\right]\right\},
\end{align}
and
\begin{align}
\Gamma_{yy}^{\mathrm{IB}}(\omega)
=&\int_{-1}^{1}\frac{2\sqrt{1-x^2} dx}{\pi}
\left\{\Theta\left[-x+\xi_{+}\right]-\Theta\left[-x+\xi_{-}\right]\right\}.
\end{align}
with $\xi_{\pm}=\frac{2\mu \pm\omega}{\omega}\frac{\Theta(t)}{t}$.

By imposing constraints on the integration interval via the Heaviside step function, $\Gamma_{jj}^{\mathrm{IB}}(\omega)$ can be obtained for the type-I, type-II, and type-III Dirac bands as follows.

\subsubsection{Interband LOC for type-I phase}

For the type-I phase ($0<t<1$), due to $\xi_{+}=(2\mu/\omega+1)\frac{\Theta(t)}{t}>1$, we have $\Theta\left[-x+\xi_{+}\right]=1$. The energy $\omega$ have three region determined by the Heaviside step function $\Theta\left[-x+\xi_{-}\right]$, which are $0<\omega<\frac{2\mu}{t+1}$ for $\Theta\left[-x+\xi_{-}\right]=1$ at $-1<x<1$, $\frac{2\mu}{t+1}\leq\omega<\frac{2\mu}{t-1}$ for $\Theta\left[-x+\xi_{-}\right]=1$ at $-1<x<\xi_-$, and $\omega\geq\frac{2\mu}{t-1}$ for $\Theta\left[-x+\xi_{-}\right]=0$ at $-1<x<1$. The $\Gamma_{jj}^{\mathrm{IB}}(\omega)$ are given by
\begin{align}
\Gamma_{xx}^{\mathrm{IB}}(\omega)
=1-\int_{-1}^{1}\frac{2x^2 dx}{\pi\sqrt{1-x^2}}\Theta\left[-x+\xi_{-}\right]
=\begin{cases}
0, & 0<\omega<\omega_1(t)\\\\
1-G_{-}(\xi_{-}), & \omega_1(t)\leq\omega<\omega_2(t)\\\\
1, & \omega\geq\omega_2(t),
\end{cases}
\end{align}
and
\begin{align}
\Gamma_{yy}^{\mathrm{IB}}(\omega)
=1-\int_{-1}^{1}\frac{2\sqrt{1-x^2} dx}{\pi}\Theta\left[-x+\xi_{-}\right]
=\begin{cases}
0, & 0<\omega<\omega_1(t)\\\\
1-G_{+}(\xi_{-}), & \omega_1(t)\leq\omega<\omega_2(t)\\\\
1, & \omega\geq\omega_2(t),
\end{cases}
\end{align}
where
\begin{align}
\omega_1(t)=&2\mu \frac{\Theta(t)}{1+t}, \notag\\
\omega_2(t)
=&2\mu \Theta(t)\left[\frac{\Theta(1-t)}{1-t}+\frac{\Theta(t-1)}{t-1}\right],\notag\\
G_\pm(x)=&\frac{1}{2}+\frac{\arcsin x}{\pi}\pm\frac{x\sqrt{1-x^2}}{\pi}.
\end{align}

\subsubsection{Interband LOC for type-II phase}

At the type-II phase ($t>1$), the Heaviside step function $\Theta\left[-x+\xi_{+}\right]$ is equal to 1, when $0<\omega<\frac{2\mu}{t-1}$ for $x\in(-1,1)$, and when $\omega\geq\frac{2\mu}{t-1}$ for $x\in(-1,\xi_+)$. The second Heaviside step function satisfies $\Theta\left[-x+\xi_{-}\right]=1$ when $0<\omega<\frac{2\mu}{t+1}$ for $x\in(-1,1)$, and when $\omega\geq\frac{2\mu}{t+1}$ for $x\in(-1,\xi_-)$ so that there are three regions of the energy $\omega$, which are $0<\omega<\frac{2\mu}{t+1}$, $\frac{2\mu}{t+1}\leq\omega<\frac{2\mu}{t-1}$, and $\omega\geq\frac{2\mu}{t-1}$. So, we have
\begin{align}
\Gamma_{xx}^{\mathrm{IB}}(\omega)
=\int_{-1}^{1}\frac{2x^2 dx}{\pi\sqrt{1-x^2}}
\left\{\Theta\left[-x+\xi_{+}\right]-\Theta\left[-x+\xi_{-}\right]\right\}
=\begin{cases}
0, & 0<\omega<\omega_1(t)\\\\
1-G_{-}(\xi_{-}), & \omega_1(t)\leq\omega<\omega_2(t)\\\\
\sum\limits_{\chi=\pm1}\chi G_{-}(\xi_{\chi}), & \omega\geq\omega_2(t)
\end{cases}
\end{align}
and
\begin{align}
\Gamma_{yy}^{\mathrm{IB}}(\omega)
=\int_{-1}^{1}\frac{2\sqrt{1-x^2} dx}{\pi}
\left\{\Theta\left[-x+\xi_{+}\right]-\Theta\left[-x+\xi_{-}\right]\right\}
=\begin{cases}
0, & 0<\omega<\omega_1(t)\\\\
1-G_{+}(\xi_{-}), & \omega_1(t)\leq\omega<\omega_2(t)\\\\
\sum\limits_{\chi=\pm1}\chi G_{+}(\xi_{\chi}), & \omega\geq\omega_2(t).
\end{cases}
\end{align}

\subsubsection{Interband LOC for type-III phase}

For the type-III phase ($t=1$), due to $\xi_{+}=2\mu/\omega+1>1$, we have $\Theta\left[-x+\xi_{+}\right]=1$, and now the $\xi_{-}$ is $\xi_{-}=2\mu/\omega-1$. The energy $\omega$ has two regions by solving the Heaviside step function $\Theta\left[-x+\xi_{-}\right]=1$ or $0$, which are $0<\omega<\mu$ and $\omega\geq\mu$. The $\Gamma_{jj}^{\mathrm{IB}}(\omega)$ are given by
\begin{align}
\Gamma_{xx}^{\mathrm{IB}}(\omega)
=1-\int_{-1}^{1}\frac{2x^2 dx}{\pi\sqrt{1-x^2}}\Theta\left[-x+\xi_{-}\right]
=\begin{cases}
0, & 0<\omega<\mu\\\\
1-G_{-}(\xi_{-}), & \omega\geq\mu
\end{cases}
\end{align}
and
\begin{align}
\Gamma_{yy}^{\mathrm{IB}}(\omega)
=1-\int_{-1}^{1}\frac{2\sqrt{1-x^2} dx}{\pi}\Theta\left[-x+\xi_{-}\right]
=\begin{cases}
0, & 0<\omega<\mu\\\\
1-G_{+}(\xi_{-}), & \omega\geq\mu.
\end{cases}
\end{align}

\subsection{Detailed calculation of intraband LOC}
After substituting $\tilde{k}_x=v_x k_x$ and $\tilde{k}_y=v_y k_y$ into Eq. (\ref{Eq6}) and expressing it in terms of $\tilde{k}=\sqrt{\tilde{k}_x^2+\tilde{k}_y^2}$ and $\phi=\arctan(\tilde{k}_y/\tilde{k}_x)$, the original expressions in the anisotropic model are converted to be the rescaled forms in the isotropic model as
\begin{align}
\mathrm{Re}\sigma_{jj(\mathrm{D})}^{\kappa,\lambda}(\omega)
=\int \frac{d\tilde{k}_xd\tilde{k}_y}{4\pi v_xv_y}\mathcal{\tilde{F}}_{\lambda,\lambda}^{\kappa;jj} (\tilde{k}_x,\tilde{k}_y) \left[-\frac{df[\tilde{\varepsilon}_\kappa^\lambda(\tilde{k}_x,\tilde{k}_y)]} {d\tilde{\varepsilon}_\kappa^\lambda(\tilde{k}_x,\tilde{k}_y)}\right] \delta(\omega),
\end{align}
with
\begin{align}
\mathcal{\tilde{F}}_{\lambda,\lambda}^{\kappa;xx}(\tilde{k}_x,\tilde{k}_y)&=
4\sigma_0\frac{v_x^2\tilde{k}_x^2}{\tilde{k}^2} =4\sigma_0v_x^2\cos^2\phi,\notag\\
\mathcal{\tilde{F}}_{\lambda,\lambda}^{\kappa;yy}(\tilde{k}_x,\tilde{k}_y) &=
4\sigma_0v_y^2\left(t ^2+\frac{\tilde{k}_y^2}{\tilde{k}^2} +2\frac{\lambda\kappa t \tilde{k}_y}{\tilde{k}}\right) =4\sigma_0v_y^2(\sin\phi+\lambda\kappa t)^2,\notag
\end{align}
where $\tilde{\varepsilon}_\kappa^\lambda(\tilde{k}_x,\tilde{k}_y)=\kappa t \tilde{k}_y+\lambda\tilde{k}$.

At zero temperature, the Fermi distribution function $f(x)$ can be replaced by the Heaviside step function
$\Theta[\mu -x]$, then the derivative reduces to $\delta\left[\mu-\tilde{\varepsilon}_\kappa^\lambda(\tilde{k}_x,\tilde{k}_y)\right]$. In the polar coordinate, we obtain
\begin{align}
&\mathrm{Re}\sigma_{xx(\mathrm{D})}^{\kappa,\lambda}(\omega)
=\frac{\sigma_0}{\pi}\frac{v_x}{v_y}\int^{+\infty}_{0}\tilde{k}d\tilde{k} \int^{3\pi/2}_{-\pi/2}\cos^2\phi d\phi \delta\left[\mu-(\kappa t \sin\phi+\lambda)\tilde{k}\right]\delta(\omega)\notag\\
&=\frac{\sigma_0}{\pi}\frac{v_x}{v_y}\int^{+\infty}_{0}\tilde{k}d\tilde{k} \int^{\pi/2}_{-\pi/2}\cos^2\phi d\phi \left\{\delta\left[\mu-(\kappa t \sin\phi+\lambda)\tilde{k}\right] +\delta\left[\mu-(-\kappa t \sin\phi+\lambda)\tilde{k}\right]\right\}\delta(\omega),\\
&\mathrm{Re}\sigma_{yy(\mathrm{D})}^{\kappa,\lambda}(\omega)
=\frac{\sigma_0}{\pi}\frac{v_y}{v_x}\int^{+\infty}_{0}\tilde{k}d\tilde{k} \int^{3\pi/2}_{-\pi/2}(\sin\phi+\lambda\kappa t)^2 d\phi \delta\left[\mu-(\kappa t \sin\phi+\lambda)\tilde{k}\right]\delta(\omega)\notag\\
&=\frac{\sigma_0}{\pi}\frac{v_y}{v_x}\int^{+\infty}_{0}\tilde{k}d\tilde{k} \int^{\pi/2}_{-\pi/2}(\sin\phi+\lambda\kappa t)^2 d\phi \left\{\delta\left[\mu-(\kappa t \sin\phi+\lambda)\tilde{k}\right] +\delta\left[\mu-(-\kappa t \sin\phi+\lambda)\tilde{k}\right]\right\}\}\delta(\omega).
\end{align}

As a consequence, the ratio $v_y/v_x$ or $v_x/v_y$ can be factored out from the original expressions as
\begin{align}
\mathrm{Re}\sigma_{xx(\mathrm{D})}^{\lambda}(\omega)=&g_s\left[\mathrm{Re}\sigma_{xx(\mathrm{D})}^{+,\lambda}(\omega) +\mathrm{Re}\sigma_{xx(\mathrm{D})}^{-,\lambda}(\omega)\right]
=\sigma_0\frac{v_x}{v_y}\Gamma_{xx}^{\mathrm{D},\lambda}(\mu,t,\Lambda)\delta(\omega),\\
\mathrm{Re}\sigma_{yy(\mathrm{D})}^{\lambda}(\omega) =&g_s\left[\mathrm{Re}\sigma_{yy(\mathrm{D})}^{+,\lambda}(\omega) +\mathrm{Re}\sigma_{yy(\mathrm{D})}^{-,\lambda}(\omega)\right]
=\sigma_0\frac{v_y}{v_x}\Gamma_{yy}^{\mathrm{D},\lambda}(\mu,t,\Lambda)\delta(\omega),
\end{align}
where two dimensionless auxiliary functions
\begin{align}
\Gamma_{xx}^{\mathrm{D},\lambda}(\mu,t,\Lambda)=&
\frac{8}{\pi}\int^{+\infty}_{0}\tilde{k}d\tilde{k}
\int^{\pi/2}_{-\pi/2}\cos^2\phi d\phi~\delta\left[\mu-(t\sin\phi+\lambda)\tilde{k}\right],\\
\Gamma_{yy}^{\mathrm{D},\lambda}(\mu,t,\Lambda)=&
\frac{8}{\pi}\int^{+\infty}_{0}\tilde{k}d\tilde{k}
\int^{\pi/2}_{-\pi/2} (\sin\phi+\lambda t)^2 d\phi~\delta\left[\mu-(t\sin\phi+\lambda)\tilde{k}\right],
\end{align}
are introduced for convenience. The ratio $v_x/v_y$ in $\mathrm{Re}\sigma_{xx(\mathrm{D})}^{\lambda}(\omega)$ and $v_y/v_x$ in $\mathrm{Re}\sigma_{yy(\mathrm{D})}^{\lambda}(\omega)$ are totally different in the anisotropic model but the same in the isotropic model. Therefore, the calculation of $\mathrm{Re}\sigma_{xx(\mathrm{D})}^{\lambda}(\omega)$ and $v_y/v_x$ in $\mathrm{Re}\sigma_{yy(\mathrm{D})}^{\lambda}(\omega)$ boils down to calculating $\Gamma_{xx}^{\mathrm{D},\lambda}(\mu,t,\Lambda)$ and $\Gamma_{yy}^{\mathrm{D},\lambda}(\mu,t,\Lambda)$.

\subsubsection{Intraband LOC for type-I phase}
For the type-I phase ($0<t<1$), by replacing $\sin\phi$ with $x$ and integrating over $\tilde{k}$, we get
%\begin{align}
%\Gamma_{xx}^{\mathrm{D},\mathrm{sgn}(\mu)}(\omega)
%&=\mathcal{N}_{\mathrm{I},xx}^{\mathrm{D},\mathrm{sgn}(\mu)}(t)\delta(\omega),\\
%\Gamma_{yy}^{\mathrm{D},\mathrm{sgn}(\mu)}(\omega)
%&=\mathcal{N}_{\mathrm{I},yy}^{\mathrm{D},\mathrm{sgn}(\mu)}(t)\delta(\omega).
%\end{align}
%where
\begin{align}
\Gamma_{xx}^{\mathrm{D},\mathrm{sgn}(\mu)}(\mu,t,\Lambda)
&=\frac{8\mu}{\pi}\int^{1}_{-1}dx\frac{\sqrt{1-x^2}}{[1+\mathrm{sgn}(\mu)t x]^2}
=8\mu \frac{1-\sqrt{1-t^2}}{t^2\sqrt{1-t^2}},\\
\Gamma_{yy}^{\mathrm{D},\mathrm{sgn}(\mu)}(\mu,t,\Lambda)
&=\frac{8\mu}{\pi}\int^{1}_{-1}\frac{dx}{[1+\mathrm{sgn}(\mu)t x]^2}\frac{[x+\mathrm{sgn}(\mu)t]^2}{\sqrt{1-x^2}}=8\mu \frac{1-\sqrt{1-t^2}}{t^2}.
\end{align}

Utilizing the relations
\begin{align}
\mathrm{Re}\sigma_{xx(\mathrm{D})}^{\lambda}(\omega)&=\sigma_0\frac{v_x}{v_y}
\Gamma_{xx}^{\mathrm{D},\mathrm{sgn}(\mu)}(\mu,t,\Lambda)\delta(\omega)
\equiv \sigma_0\frac{4\mu}{\pi} N_1\delta(\omega),\\
\mathrm{Re}\sigma_{yy(\mathrm{D})}^{\lambda}(\omega)&=\sigma_0\frac{v_y}{v_x}
\Gamma_{yy}^{\mathrm{D},\mathrm{sgn}(\mu)}(\mu,t,\Lambda)\delta(\omega)
\equiv \sigma_0\frac{4\mu}{\pi} N_2\delta(\omega),
\end{align}
we have
\begin{align}
N_1&=2\pi\frac{v_x}{v_y}\frac{1-\sqrt{1-t^2}}{t^2\sqrt{1-t^2}},\\
N_2&=2\pi\frac{v_y}{v_x}\frac{1-\sqrt{1-t^2}}{t^2}.
\end{align}
These expressions of Drude conductivity yield the numerical results of Drude weight in the Ref.\cite{PRBVerma2017}, namely, $N_1=4.686$ and $N_2=2.673$ after substituting the parameters of 8-$Pmmn$ borophene ($v_x=0.86v_F$, $v_y=0.69v_F$ and $v_t=0.32v_F$ with $v_F=10^6m/s$).

In the limit $t\to0$, we recover the result for ordinary Dirac cone,
\begin{align}
\Gamma_{xx}^{\mathrm{D},\mathrm{sgn}(\mu)}(\mu,t=0,\Lambda)
=\Gamma_{yy}^{\mathrm{D},\mathrm{sgn}(\mu)}(\mu,t=0,\Lambda)=4\mu=4\mu~\mathrm{sgn}(\mu).
\end{align}

In the limit $t\to1^{-}$, we have
\begin{align}
\Gamma_{xx}^{\mathrm{D},\mathrm{sgn}(\mu)}(\mu,t=1^{-},\Lambda)&\to\infty,\\
\Gamma_{yy}^{\mathrm{D},\mathrm{sgn}(\mu)}(\mu,t=1^{-},\Lambda)&=8\mu=8\mu~\mathrm{sgn}(\mu).
\end{align}

\subsubsection{Intraband LOC for type-II phase}
For the type-II phase ($t>1$), there is not only an electron pocket but also a hole pocket at one valley. We take $\kappa=+1$ and $\mu>0$ as an example. In the polar coordinates, there are constraints on the values of $\tilde{k}$ and $\phi$. For $\lambda=+1$, $\tilde{k}_{min}=\mu/(t+1)$, $\phi\in(\phi_1,\pi+|\phi_1|)$ where $\phi_1$ is determined by $(t\sin\phi_1+1)\Lambda=\mu$, where $\Lambda$ is the cutoff of $\tilde{k}$. The cut-off $\Lambda$ is a measure of the density of states due to electron and hole Fermi pockets \cite{PRBCarbotte2016}. For $\lambda=-1$, $\tilde{k}_{min}=\mu/(t-1)$, $\phi\in(\phi_2,\pi-\phi_2)$, where $\phi_2$ is obtained by solving $(t\sin\phi_2-1)\Lambda=\mu$. The schematic diagrams of these constrains in the polar coordinate are explicitly shown in Fig. \ref{fig5}.

\begin{figure}[htbp]
\centering
\includegraphics[width=16cm]{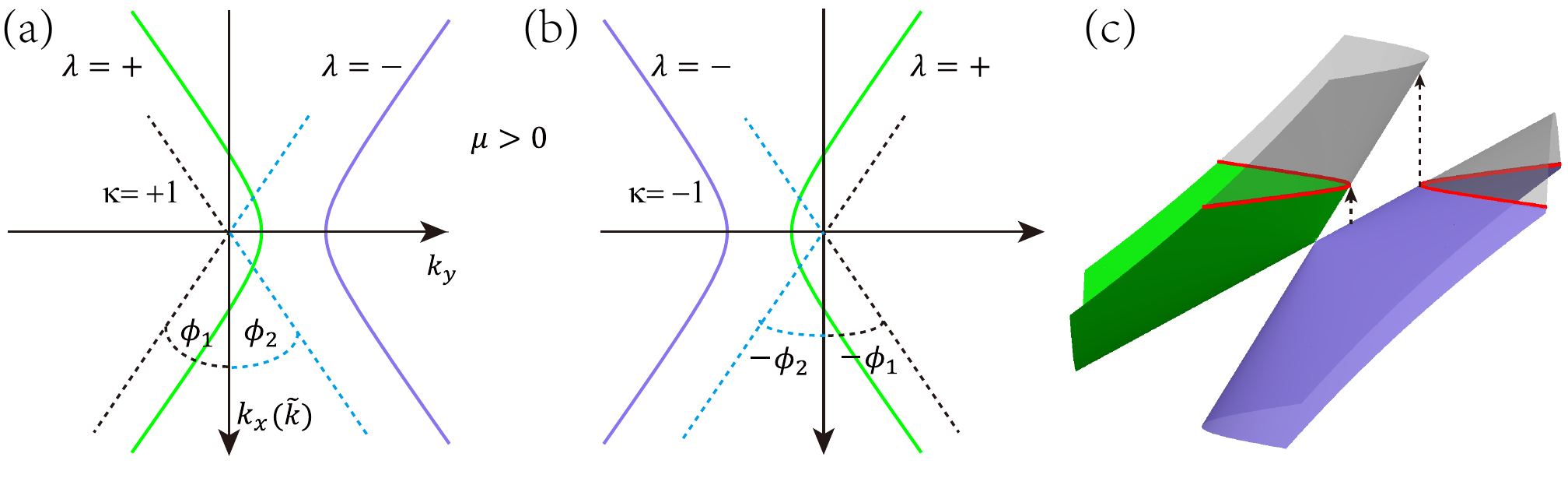}
\caption{Schematic diagram of the limitation of the angle $\phi$ imposed by the delta function in the $n$-doped case ($\mu>0$). In
the $\kappa=+1$ valley shown in (a), the angle ranges from $\phi_1$ to $\pi+|\phi_1|$ for $\lambda=+$, but becomes ($\phi_2$, $\pi-\phi_2$) for $\lambda=-$. The case for $\kappa=-$ is shown in (b) for reference. The corresponding optical transitions at the $\kappa=+$ valley are referred to in (c).}
\label{fig5}
\end{figure}

Consequently, for $\lambda=+1$, $\Gamma_{xx}^{\mathrm{D},+}(\mu,t,\Lambda)$ is given by
\begin{align}
\Gamma_{xx}^{\mathrm{D},+}(\mu,t,\Lambda)
&=\frac{8}{\pi}\int^{\Lambda}_{\mu/(1+t)}\tilde{k}d\tilde{k}\int^{\pi/2}_{\phi_1}\cos^2\phi d\phi \delta\left[\mu-(t \sin\phi+1)\tilde{k}\right]\notag\\
&=\frac{8}{\pi}\int^{\Lambda}_{\mu/(t+1)}\tilde{k}d\tilde{k} \int^{1}_{[(\mu/\Lambda)-1](1/t)}\sqrt{1-x^2} dx \delta\left[\mu-(tx+1)\tilde{k}\right]\notag\\
&=\frac{8\mu}{\pi}\int^{1}_{[(\mu/\Lambda)-1](1/t)} \frac{\sqrt{1-x^2}}{(t x+1)^2} dx\Theta\left[\Lambda-\frac{\mu}{t+1}\right].
\end{align}

Similarly, for $\lambda=-1$, $\Gamma_{xx}^{\mathrm{D},-}(\mu,t,\Lambda)$ reads
\begin{align}
\Gamma_{xx}^{\mathrm{D},-}(\mu,t,\Lambda)
=\frac{8\mu}{\pi}\int^{1}_{[(\mu/\Lambda)+1](1/t)} \frac{\sqrt{1-x^2}}{(tx-1)^2} dx\Theta\left[\Lambda-\frac{\mu}{t-1}\right].
\end{align}

As a consequence, $\Gamma_{xx}^{\mathrm{D}}(\mu,t,\Lambda)$ can be written as
\begin{align}
\Gamma_{xx}^{\mathrm{D}}(\mu,t,\Lambda)=&\Gamma_{xx}^{\mathrm{D},-}(\mu,t,\Lambda)+\Gamma_{xx}^{\mathrm{D},+}(\mu,t,\Lambda)\notag\\
&=\frac{8\mu }{\pi}\left\{\int^{1}_{[(\mu/\Lambda)-1][(1/t)} \frac{\sqrt{1-x^2}}{(t x+1)^2} dx\Theta\left[\Lambda-\frac{\mu }{t+1}\right]+\int^{1}_{[(\mu/\Lambda)+1][(1/t)} \frac{\sqrt{1-x^2}}{(t x-1)^2} dx\Theta\left[\Lambda-\frac{\mu}{t-1}\right]\right\}\notag\\
&=\frac{8\mu}{\pi}\left[A(\mu ,t,\Lambda)\frac{2\Lambda}{\mu}+\frac{B(\mu,t,\Lambda)}{\sqrt{t^2-1}}-C(\mu ,t,\Lambda)\right],
\end{align}
where
\begin{align}
A(\mu ,t,\Lambda)=&\sum_{\chi=\pm}\frac{1}{2t}\sqrt{1-\left(\frac{\mu -\chi\Lambda}{t\Lambda}\right)^2},\notag\\
B(\mu ,t,\Lambda)=&\frac{1}{t^2}\ln\frac{t^2+\frac{\mu -\Lambda}{\Lambda}
+\sqrt{t^2-1}\sqrt{t^2-(\frac{\mu -\Lambda}{\Lambda})^2}} {t^2-\frac{\mu +\Lambda}{\Lambda}+\sqrt{t^2-1}\sqrt{t^2-(\frac{\mu +\Lambda}{\Lambda})^2}},\notag\\
C(\mu ,t,\Lambda)=&\sum_{\chi=\pm}\frac{1}{t^2}\arccos\frac{\mu -\chi\Lambda}{t\Lambda}.
\end{align}

Similarly, $\Gamma_{yy}^{\mathrm{D}}(\mu,t,\Lambda)$ can be written as
\begin{align}
\Gamma_{yy}^{\mathrm{D}}(\mu,t,\Lambda)&
=\Gamma_{yy}^{\mathrm{D,+}}(\mu,t,\Lambda)+\Gamma_{yy}^{\mathrm{D,-}}(\mu,t,\Lambda)
\notag\\&
=\frac{8\mu}{\pi} \left\{\int^{1}_{[(\mu/\Lambda)-1][(1/t)}
\frac{1}{(1+t x)^2}\frac{(x+t )^2}{\sqrt{1-x^2}} dx \Theta\left[\Lambda-\frac{\mu}{1+t}\right]
+\int^{1}_{[(\mu/\Lambda)+1][(1/t)} \frac{1}{(tx-1)^2} \frac{(x-t)^2}{\sqrt{1-x^2}} dx \Theta\left[\Lambda-\frac{\mu}{t-1}\right]\right\}
\notag\\&
=\frac{8\mu}{\pi} \left[(t^2-1)A(\mu ,t,\Lambda)\frac{2\Lambda}{\mu}
+\sqrt{t^2-1}B(\mu ,t,\Lambda)+C(\mu ,t,\Lambda)\right].
\end{align}

Keeping the order of $O(1)$ of $\Lambda$, we have
\begin{align}
\Gamma_{xx}^{\mathrm{D}}(\mu,t,\Lambda)
&=8\mu\left[\frac{\sqrt{t^2-1}}{\pi t^2}\frac{2\Lambda}{\mu}-\frac{1}{t^2}\right],\\
\Gamma_{yy}^{\mathrm{D}}(\mu,t,\Lambda)
&=8\mu\left[\frac{\sqrt{(t^2-1)^3}}{\pi t^2}\frac{2\Lambda}{\mu}+\frac{1}{t^2}\right].
\end{align}

\subsubsection{Intraband LOC for type-III phase}

For the type-III phase ($t=1$), $\Gamma_{xx}^{\mathrm{D},\mathrm{sgn}(\mu)}(\mu,t,\Lambda)$ and $\Gamma_{yy}^{\mathrm{D},\mathrm{sgn}(\mu)}(\mu,t,\Lambda)$ can be given by
\begin{align}
\Gamma_{xx}^{\mathrm{D},\mathrm{sgn}(\mu)}(\mu,t,\Lambda)
&=\frac{8\mu }{\pi}\int^{1}_{(\mu/\Lambda)-1} \frac{\sqrt{1-x^2}}{(1+x)^2} dx\Theta\left[\Lambda-\frac{\mu }{2}\right]
=\frac{16\mu}{\pi}\left[\sqrt{\frac{2\Lambda}{\mu}-1}-\arccos\sqrt{\frac{\mu}{2\Lambda}}\right]
\notag\\&
=\frac{16\mu}{\pi}\left[\sqrt{\frac{2\Lambda-\mu }{\mu }}-\arccos\sqrt{\frac{\mu}{2\Lambda}}\right],\\
\Gamma_{yy}^{\mathrm{D},\mathrm{sgn}(\mu)}(\mu,t,\Lambda)
&=\frac{8\mu}{\pi}\int^{1}_{(\mu/\Lambda)-1} \frac{1}{(1+x)^2} \frac{(x+1)^2}{\sqrt{1-x^2}} dx\Theta\left[\Lambda-\frac{\mu}{2}\right]
=\frac{8\mu}{\pi}\arccos\left(\frac{\mu}{\Lambda}-1\right)
\notag\\&
=\frac{8\mu}{\pi}\arccos\left(\frac{\mu -\Lambda}{\Lambda}\right).
\end{align}

Keeping the order of $O(1)$ of $\Lambda$, we have
\begin{align}
\Gamma_{xx}^{\mathrm{D},\mathrm{sgn}(\mu)}(\mu,t,\Lambda)
&=\frac{16\mu}{\pi}\left[\sqrt{\frac{2\Lambda-\mu }{\mu }}-\arccos\sqrt{\frac{\mu}{2\Lambda}}\right]
=8\mu\left[\frac{2}{\pi}\sqrt{\frac{2\Lambda}{\mu}}-1\right],
\end{align}
and
\begin{align}
\Gamma_{yy}^{\mathrm{D},\mathrm{sgn}(\mu)}(\mu,t,\Lambda)
&=\frac{8\mu }{\pi}\arccos\left(\frac{\mu -\Lambda}{\Lambda}\right)
=8\mu.
\end{align}

\section{Detailed calculation of JDOS \label{Sec:AppendixC}}

In the following, we present the detailed calculation of JDOS in the $n$-doped case [$\mu>0$, namely, $\mathrm{sgn}(\mu)=+$] for the type-I, type-II, and type-III Dirac bands.

\subsection{Calculation of JDOS for type-I phase}

For the type-I phase ($0<t<1$), the conduction band is partially occupied
by electrons, and the Fermi wave vectors are $\tilde{k}_F^{\kappa,+}(\mu)=\mu/(\kappa t \sin\phi+1)$. Due to the Pauli blocking, the energy of the photon must excite electrons from the valence band to the conduction band above the Fermi surface, which requires $\omega\geq \tilde{\varepsilon}_\kappa^+\left[\tilde{k}_F^{\kappa,+}(\mu)\right]- \tilde{\varepsilon}_\kappa^-\left[\tilde{k}_F^{\kappa,+}(\mu)\right]=2 \tilde{k}_F^{\kappa,+}(\mu) =2\mu /(\kappa t\sin\phi+1)$  with $\phi\in[0,2\pi]$. The JDOS at the $+\kappa$ valley can be written as
\begin{align}
\mathcal{J}_{\kappa}(\omega)&=\frac{\mathcal{J}_0}{2\pi} \int_0^{+\infty}\tilde{k}d\tilde{k}
\int_{0}^{2\pi}d\phi \delta(2\tilde{k}-\omega) \Theta\left[\omega-2 \tilde{k}_F^{\kappa,+}(\mu)\right]\notag\\
&=\frac{\mathcal{J}_0}{2\pi}\int_0^{+\infty}\tilde{k}d\tilde{k} \int_{0}^{2\pi}d\phi \delta(2\tilde{k}-\omega) \Theta\left[\omega-\frac{2\mu }{\kappa t\sin\phi+1}\right]\notag\\
&=\mathcal{J}_0 \frac{\omega}{8\pi} \int_{-\pi/2}^{3\pi/2}d\phi
\Theta\left[\omega-\frac{2\mu }{\kappa t\sin\phi+1}\right]\notag\\
&=\mathcal{J}_0 \frac{\omega}{8\pi} \int_{-\pi/2}^{\pi/2}d\phi \left\{\Theta\left[\omega-\frac{2\mu }{\kappa t\sin\phi+1}\right]
+\Theta\left[\omega-\frac{2\mu }{-\kappa t\sin\phi+1}\right]\right\}=\mathcal{J}_{-\kappa}(\omega).
\end{align}
As a result, $\mathcal{J}(\omega)=g_sg_v\mathcal{J}_{\kappa}(\omega)$, where $g_v=2$ denotes the valley degeneracy. After introducing $x=\sin\phi$, one can obtain
\begin{align}
\mathcal{J}(\omega)&=g_sg_v\mathcal{J}_{\kappa}(\omega)
=\mathcal{J}_0 \frac{\omega}{2\pi} \int_{-1}^{1}\frac{dx}{\sqrt{1-x^2}} \left\{\Theta\left[\omega-\frac{2\mu }{t x+1}\right] +\Theta\left[\omega-\frac{2\mu }{-t x+1}\right]\right\}\notag\\
&=\mathcal{J}_{0}\frac{\omega}{\pi} \int_{-1}^{1}\frac{dx}{\sqrt{1-x^2}} \Theta\left[\omega-\frac{2\mu }{t x+1}\right] =\mathcal{J}_{0}\omega
\begin{cases}
0, &0<\omega<\omega_1(t)\\\\
\frac{\arccos\xi_{-}}{\pi}, &\omega_1(t)\leq\omega<\omega_2(t)\\\\
1, &\omega\geq\omega_2(t),
\end{cases}
\end{align}
where
\begin{align}
\xi_{\pm}&=\frac{2\mu \pm\omega}{\omega}\frac{\Theta(t)}{t},\\
\omega_1(t)&= 2\mu \frac{\Theta(t)}{1+t},\\
\omega_2(t)&= 2\mu \left[\frac{\Theta(1-t)}{1-t}+\frac{\Theta(t-1)}{t-1}\right].
\end{align}
In addition, it is easy to obtain for the untilted case ($t=0$) that
\begin{align}
\mathcal{J}(\omega)
=\mathcal{J}_0 \frac{\omega}{2\pi} \int_{0}^{2\pi}d\phi
\Theta\left[\omega-2\mu\right]
=\mathcal{J}_{0}\omega
\begin{cases}
0, &0<\omega<2\mu \\\\
1, &\omega\geq2\mu .
\end{cases}
\end{align}

\subsection{Calculation of JDOS for type-II phase}

For the type-II phase ($t>1$), the valence band is partially occupied by holes, and the electron transition area will be restricted by a valence band and a conduction band. In order to conveniently describe the integration area of JDOS in $\tilde{\textbf{\emph{k}}}$ space, we introduce two angle parameters $\phi_1$ and $\phi_2$, which are obtained by solving $(t\sin\phi_1+1)\Lambda=\mu$ and $(t\sin\phi_2-1)\Lambda=\mu$ respectively, where $\Lambda$ is the cutoff of $\tilde{k}$, as shown in Fig. \ref{fig5}. The photon energy contributed to the LOC is limited to $2\tilde{k}_F^{\kappa,+}(\mu)\leq\omega\leq2\tilde{k}_F^{\kappa,-}(\mu)$.

In the polar coordinate, the JDOS for the $\kappa=+$ valley and the $\kappa=-$ valley can be respectively written as
\begin{align}
&\mathcal{J}_{+}(\omega)
=\frac{\mathcal{J}_0}{2\pi} \left\{ \int_{\mu/(t+1)}^{\Lambda}\tilde{k}d\tilde{k} \int_{\phi_1}^{\pi+|\phi_1|}d\phi \delta(2\tilde{k}-\omega) \Theta[\omega-2\tilde{k}_F^{+,+}(\mu)]%\notag\\&
%%%%%%%%%%%%%%%%%%%%%%%%%%%%
-\int_{\mu/(t+1-}^{\Lambda}\tilde{k}d\tilde{k} \int_{\phi_2}^{\pi-\phi_2}d\phi \delta(2\tilde{k}-\omega) \Theta[\omega-2\tilde{k}_F^{+,-}(\mu)] \right\}\notag\\
%%%%%%%%%%%%%%%%%%%%%%%%%%%%%%%%%%%%
%%%%%%%%%%%%%%%%%%%%%%%%%%%%%%%%%%%%
&=\frac{\mathcal{J}_0}{2\pi} \left\{ \int_{\mu/(t+1)}^{\Lambda}\tilde{k}d\tilde{k} \int_{\phi_1}^{\pi+|\phi_1|}d\phi \delta(2\tilde{k}-\omega) \Theta\left[\omega-\frac{2\mu}{t \sin\phi+1}\right]%\notag\\&
%%%%%%%%%%%%%%%%%%%%%%%%%%%%
-\int_{\mu/(t-1)}^{\Lambda}\tilde{k} d\tilde{k} \int_{\phi_2}^{\pi-\phi_2}d\phi \delta(2\tilde{k}-\omega) \Theta\left[\omega-\frac{2\mu}{t\sin\phi-1}\right] \right\}\notag\\
%%%%%%%%%%%%%%%%%%%%%%%%%%%%%%%%%%%%
%%%%%%%%%%%%%%%%%%%%%%%%%%%%%%%%%%%%
&=\frac{\mathcal{J}_0}{\pi} \left\{ \int_{\mu/(t+1)}^{\Lambda}\tilde{k}d\tilde{k} \int_{\phi_1}^{\pi/2}d\phi \delta(2\tilde{k}-\omega) \Theta\left[\omega-\frac{2\mu}{t\sin\phi+1}\right]%\notag\\&
%%%%%%%%%%%%%%%%%%%%%%%%%%%%
-\int_{\mu/(t-1)}^{\Lambda}\tilde{k}d\tilde{k} \int_{\phi_2}^{\pi/2}d\phi \delta(2\tilde{k}-\omega) \Theta\left[\omega-\frac{2\mu}{t\sin\phi-1}\right] \right\},
\end{align}
and
\begin{align}
\mathcal{J}_{-}(\omega)
=\frac{\mathcal{J}_0}{\pi} & \left\{ \int_{\mu/(t+1)}^{\Lambda}\tilde{k}d\tilde{k}
\int_{-\pi/2}^{-\phi_1}d\phi \delta(2\tilde{k}-\omega) \Theta\left[\omega-\frac{2\mu}{-t\sin\phi+1}\right]
-\int_{\mu/(t-1)}^{\Lambda}\tilde{k}d\tilde{k} \int_{-\pi/2}^{-\phi_2}d\phi \delta(2\tilde{k}-\omega) \Theta\left[\omega-\frac{2\mu}{-t\sin\phi-1}\right] \right\},
\end{align}
where $\Theta[\Lambda-\frac{\mu}{t+1}]$ and $\Theta[\Lambda-\frac{\mu}{t-1}]$ are omitted here.

By replacing $\sin \phi$ with $x$ and integrating over $\tilde{k}$, we get
\begin{align}
\mathcal{J}(\omega)=&g_s\left[\mathcal{J}_{+}(\omega)+\mathcal{J}_{-}(\omega)\right]\notag\\
=&g_s\left[\mathcal{J}_0\frac{\omega}{4\pi}\left\{ \int_{[(\mu/\Lambda)-1](1/t)}^{1} \frac{dx}{\sqrt{1-x^2}} \Theta\left[\omega-\frac{2\mu}{|t x+1|}\right]
-\int_{[(\mu/\Lambda)+1](1/t)}^{1} \frac{dx}{\sqrt{1-x^2}} \Theta\left[\omega-\frac{2\mu}{|t x-1|}\right]\right\}\right.\notag\\
%%%%%%%%%%%%%%%%%%%%%%%%%%%%%%
&\left.+\mathcal{J}_0\frac{\omega}{4\pi}\left\{ \int^{-[(\mu/\Lambda)-1](1/t)}_{-1} \frac{dx}{\sqrt{1-x^2}} \Theta\left[\omega-\frac{2\mu}{-t x+1|}\right]
-\int^{-[(\mu/\Lambda)+1](1/t)}_{-1} \frac{dx}{\sqrt{1-x^2}} \Theta\left[\omega-\frac{2\mu}{-t x-1}\right]\right\}\right]\notag\\
%%%%%%%%%%%%%%%%%%%%%%%%%%%%%%
%%%%%%%%%%%%%%%%%%%%%%%%%%%%%%
=&\mathcal{J}_0\frac{\omega}{\pi}\left\{ \int_{[(\mu/\Lambda)-1](1/t)}^{1} \frac{dx}{\sqrt{1-x^2}} \Theta\left[\omega-\frac{2\mu}{t x+1}\right]
-\int_{[(\mu/\Lambda)+1](1/t)}^{1} \frac{dx}{\sqrt{1-x^2}} \Theta\left[\omega-\frac{2\mu}{t x-1}\right]\right\}\notag\\
=&\mathcal{J}_{0}\frac{\omega}{\pi}
\begin{cases}
0, &\omega<\omega_1(t)\\\\
\arccos\xi_{-}, &\omega_1(t)\leq\omega<\omega_2(t)\\\\
\arcsin\xi_{+}-\arcsin\xi_{-}, &\omega\geq\omega_2(t).
\end{cases}
\end{align}

\subsection{Calculation of JDOS for type-III phase}
For the type-III phase ($t=1$), we can easily obtain that
\begin{align}
\mathcal{J}(\omega)
=&\mathcal{J}_0 \frac{\omega}{2\pi} \int_{(\mu/\Lambda)-1}^{1}\frac{dx}{\sqrt{1-x^2}} \Theta\left[\omega-\frac{2\mu}{x+1}\right]
+\mathcal{J}_0 \frac{\omega}{2\pi} \int_{-1}^{-[(\mu/\Lambda)-1]}\frac{dx}{\sqrt{1-x^2}} \Theta\left[\omega-\frac{2\mu}{-x+1}\right] \notag\\
=&\mathcal{J}_0 \frac{\omega}{\pi} \int_{(\mu/\Lambda)-1}^{1}\frac{dx}{\sqrt{1-x^2}} \Theta\left[\omega-\frac{2\mu}{x+1}\right]\notag\\
=&\mathcal{J}_{0}\frac{\omega}{\pi}
\begin{cases}
0, &\omega<\mu\\\\
\arccos\xi_{-}, &\omega\geq\mu.
\end{cases}
\end{align}

\section{The relationship between JDOS and the interband LOCs \label{Sec:AppendixD}}
In this appendix we will give the connection between the interband part of the optical conductivity $\mathrm{Re}\sigma_{jj}^{\mathrm{IB}}(\omega)$ and the JDOS. For the convenience of the following elaboration, we introduce a temporary auxiliary function $\mathcal{J}_{L}(\omega)$. The JDOS can be written as $\mathcal{J}(\omega)=(\omega\mathcal{J}_0)\times\mathcal{J}_{L}(\omega)$.

The complete formalisms for $\mathrm{Re}\sigma_{xx}^\kappa(\omega)$ and $\mathrm{Re}\sigma_{yy}^\kappa(\omega)$ are as follows (at zero temperature $T\to0$):
\begin{align}
\mathrm{Re}\sigma_{xx}^{\kappa(\mathrm{IB})}(\omega) &= 2\sigma_0v_x^2\pi \int\frac{d^2\boldsymbol{k}}{(2\pi)^2}
 \left\{1-\frac{v_x^2k_x^2-v_y^2k_y^2} {\left[\mathcal{Z}(k_x,k_y)\right]^2}\right\} \frac{\Theta\left[\mu-\varepsilon^-_{\kappa}(k_x,k_y)\right] -\Theta\left[\mu-\varepsilon_\kappa^{+}(k_x,k_y)\right]}{\omega} \delta\left[\omega-2\mathcal{Z}(k_x,k_y)\right]\notag\\
&= 2\sigma_0v_x^2\pi \left\{ \frac{\mathcal{J}_{\kappa}(\omega)}{\omega}-\int\frac{d^2\boldsymbol{k}}{(2\pi)^2} \frac{v_x^2k_x^2-v_y^2k_y^2} {\left[\mathcal{Z}(k_x,k_y)\right]^2} \frac{\Theta\left[\mu-\varepsilon^-_{\kappa}(k_x,k_y)\right] -\Theta\left[\mu-\varepsilon_\kappa^{+}(k_x,k_y)\right]}{\omega} \delta\left[\omega-2\mathcal{Z}(k_x,k_y)\right]\right\}\notag\\
&= \sigma_0 \left[2v_x^2\pi \frac{\mathcal{J}_{\kappa}(\omega)}{\omega}-\frac{v_x}{v_y}\mathcal{L}_\kappa(\omega) \right],
\end{align}
\begin{align}
\mathrm{Re}\sigma_{yy}^{\kappa(\mathrm{IB})}(\omega) &= 2\sigma_0v_y^2\pi \int\frac{d^2\boldsymbol{k}}{(2\pi)^2} \left\{1+\frac{v_x^2k_x^2-v_y^2k_y^2} {\left[\mathcal{Z}(k_x,k_y)\right]^2}\right\} \frac{\Theta\left[\mu-\varepsilon^-_{\kappa}(k_x,k_y)\right] -\Theta\left[\mu-\varepsilon_\kappa^{+}(k_x,k_y)\right]}{\omega} \delta\left[\omega-2\mathcal{Z}(k_x,k_y)\right]\notag\\
&= \sigma_0\left[2v_y^2\pi \frac{\mathcal{J}_{\kappa}(\omega)}{\omega}+\frac{v_y}{v_x}\mathcal{L}_\kappa(\omega) \right],
\end{align}
where
\begin{align}
\mathcal{L}_\kappa(\omega)&=2\pi v_xv_y \int\frac{d^2\boldsymbol{k}}{(2\pi)^2} \frac{v_x^2k_x^2-v_y^2k_y^2} {\left[\mathcal{Z}(k_x,k_y)\right]^2} \frac{\Theta\left[\mu-\varepsilon^-_{\kappa}(k_x,k_y)\right] -\Theta\left[\mu-\varepsilon_\kappa^{+}(k_x,k_y)\right]}{\omega} \delta\left[\omega-2\mathcal{Z}(k_x,k_y)\right]
\end{align}

The final formalism for $\mathrm{Re}\sigma_{jj}^{\mathrm{IB}}(\omega)$ are
\begin{align}
\mathrm{Re}\sigma_{xx}^{\mathrm{IB}}(\omega)
&=g_s\sum_{\kappa=\pm1} \mathrm{Re}\sigma_{xx}^{\kappa(\mathrm{IB})}(\omega)
=\sigma_0\left[2v_x^2\pi g_s\sum_{\kappa=\pm1}\frac{\mathcal{J}_\kappa(\omega)}{\omega}-g_s \sum\limits_{\kappa=\pm1}\frac{v_x}{v_y}\mathcal{L}_\kappa(\omega)\right]\notag\\
&=\sigma_0\left[2v_x^2\pi \frac{\mathcal{J}(\omega)}{\omega}-\frac{v_x}{v_y}\mathcal{L}(\omega)\right]
=\sigma_0\left[\frac{2v_x^2\pi }{\omega}(\omega\mathcal{J}_0)\mathcal{J}_{L}(\omega)-\frac{v_x}{v_y}\mathcal{L}(\omega)\right]
\notag\\&
=\sigma_0\left[\frac{2v_x^2\pi }{\omega}\frac{\omega}{2\pi v_xv_y}\mathcal{J}_{L}(\omega)-\frac{v_x}{v_y}\mathcal{L}(\omega)\right]
\notag\\&
=\sigma_0\frac{v_x}{v_y} \left[\mathcal{J}_{L}(\omega)-\mathcal{L}(\omega)\right] =\sigma_0\frac{v_x}{v_y} \left[\frac{\mathcal{J}(\omega)}{\mathcal{J}_0\omega}+\mathcal{L}(\omega)\right],\\
%%%%%%%%%%%%%%%%%%%%%%%%%%%%%%%%%%%%%%%%%%%%%%
\mathrm{Re}\sigma_{yy}^{\mathrm{IB}}(\omega)
&=g_s\sum_{\kappa=\pm1} \mathrm{Re}\sigma_{yy}^{\kappa(\mathrm{IB})}(\omega)
=\sigma_0\left[2v_x^2\pi \frac{\mathcal{J}(\omega)}{\omega}+\frac{v_y}{v_x}\mathcal{L}(\omega)\right]\notag\\
&=\sigma_0\frac{v_y}{v_x} \left[\mathcal{J}_{L}(\omega)+\mathcal{L}(\omega)\right] =\sigma_0\frac{v_y}{v_x} \left[\frac{\mathcal{J}(\omega)}{\mathcal{J}_0\omega}+\mathcal{L}(\omega)\right],
\end{align}
namely,
\begin{align}
\Gamma^{\mathrm{IB}}_{xx}(\omega) =\frac{\mathcal{J}(\omega)}{\mathcal{J}_0\omega}-\mathcal{L}(\omega)\notag\\
%%%%%%%%%%%%%%%%%%%%%%%%%%%%%%%%%%%%%%%%%%%%%%
\Gamma^{\mathrm{IB}}_{yy}(\omega)
=\frac{\mathcal{J}(\omega)}{\mathcal{J}_0\omega}+\mathcal{L}(\omega),
\end{align}
where
\begin{align}
\mathcal{L}(\omega)=g_s\sum\limits_{\kappa=\pm1}\mathcal{L}_\kappa(\omega).
\end{align}

After analysis and calculation similar to $\mathrm{Re}\sigma_{jj}(\omega)$, we can get
\begin{align}
\mathcal{L}(\omega)
&=\frac{1}{\pi}\int_{-1}^{1} \frac{1-2x^2}{\sqrt{1-x^2}} dx \left\{\Theta\left[-x+\xi_{+}\right] -\Theta\left[-x+\xi_{-}\right]\right\}.
\end{align}
Using the approach discussed in Appendix B for the Heaviside step functions at different tilt types, we can obtain analytical results for $\mathcal{L}(\omega)$. For type-I phase ($0<t<1$), the analytical expression
of $\mathcal{L}(\omega)$ is given as
\begin{align}
\mathcal{L}(\omega)
&=\frac{1}{\pi}
\begin{cases}
0, & 0<\omega<\omega_1(t), \\\\
-\xi_{-}\sqrt{1-\xi_{-}^2}, & \omega_1(t)\leq\omega<\omega_2(t), \\\\
0, & \omega\geq\omega_2(t).
\end{cases}
\end{align}

For the type-II phase ($t>1$), the analytical expression
of $\mathcal{L}(\omega)$ takes the form
\begin{align}
\mathcal{L}(\omega)
&=\frac{1}{\pi}
\begin{cases}
0, & 0<\omega<\omega_1(t), \\\\
-\xi_{-}\sqrt{1-\xi_{-}^2}, & \omega_1(t)\leq\omega<\omega_2(t), \\\\
\xi_{+}\sqrt{1-\xi_{+}^2}-\xi_{-}\sqrt{1-\xi_{-}^2}, & \omega\geq\omega_2(t).
\end{cases}
\end{align}
For the type-III phase ($t=1$), the auxiliary function $\mathcal{L}(\omega)$ can
be written as
\begin{align}
\mathcal{L}(\omega)
&=\frac{1}{\pi}
\begin{cases}
0, & 0<\omega<\mu, \\\\
-\xi_{-}\sqrt{1-\xi_{-}^2}, & \omega\geq\mu.
\end{cases}
\end{align}

\end{widetext}

%%The end of the Appendices%%

\end{document}